\documentclass[journal]{IEEEtran}
\usepackage{ifpdf}

\usepackage{cite}

\ifCLASSINFOpdf
  \usepackage[pdftex]{graphicx}
\else
  \usepackage[dvips]{graphicx}
\fi
\usepackage{amsmath}
\usepackage{array}
\usepackage{fixltx2e}

\usepackage{stfloats}
\usepackage{url}

\usepackage{amssymb} 
\usepackage{mathtools, nccmath}
\usepackage{algorithm}
\usepackage{algpseudocode}
\usepackage{multirow}
\usepackage{color}
\usepackage{mathrsfs}
\usepackage[table, dvipsnames]{xcolor}
\usepackage{color, colortbl}
\definecolor{LightCyan}{rgb}{0.88,1,1}
\usepackage{tikz} 
\usepackage{pgfplots}
\pgfplotsset{compat=1.5}
\usepackage{amsmath,bm}

\begin{document}
	
\markboth{Journal of \LaTeX\ Class Files,~Vol.~14, No.~8, August~2015}%
{Shell \MakeLowercase{\textit{et al.}}: Bare Demo of IEEEtran.cls for IEEE Journals}
%
	
%
\title{End-to-End Learned Image Compression with Quantized Weights and Activations}
%
%
%

\author{Heming~Sun,~\IEEEmembership{Member,~IEEE,}
        Lu~Yu,~\IEEEmembership{Member,~IEEE,}
        and~Jiro~Katto,~\IEEEmembership{Member,~IEEE}
}

%
%



\maketitle

\begin{abstract}
End-to-end Learned image compression (LIC) has reached the traditional hand-crafted methods such as BPG (HEVC intra) in terms of the coding gain. However, the large network size prohibits the usage of LIC on resource-limited embedded systems. This paper reduces the network complexity by quantizing both weights and activations. 1) For the weight quantization, we study different kinds of grouping and quantization scheme at first. A channel-wise non-linear quantization scheme is determined based on the coding gain analysis. After that, we propose a fine tuning scheme to clip the weights within a certain range so that the quantization error can be reduced. 2) For the activation quantization, we first propose multiple non-linear quantization codebooks with different maximum dynamic ranges. By selecting an optimal one through a multiplexer, the quantization range can be saturated to the greatest extent. In addition, we also exploit the mean-removed quantization for the analysis transform outputs in order to reduce the bit-width cost for the specific channel with the large non-zero mean. By quantizing each weight and activation element from 32-bit floating point to 8-bit fixed point, the memory cost for both weight and activation can be reduced by 75\% with negligible coding performance loss. As a result, our quantized LIC can still outperform BPG in terms of MS-SSIM. To our best knowledge, this is the first work to give a complete analysis on the coding gain and the memory cost for a quantized LIC network, which validates the feasibility of the hardware implementation.

\end{abstract}

\begin{IEEEkeywords}
Image compression, neural networks, quantization, fixed-point, fine tuning, specific hardware
\end{IEEEkeywords}

%
\IEEEpeerreviewmaketitle

\section{Introduction}

\IEEEPARstart{I}{mage} compression is important to relieve the burden of the image transmission and storage. In the past decades, several standards have been developed such as JPEG \cite{Wallace1992TheJS}, JPEG2000 \cite{rabbani2002overview}, WebP \cite{lian2012webp} and HEVC intra (BPG) \cite{sullivan2012overview}. For these standardized methods, the coding components are fixed which are composed of intra prediction, linear transform, quantization and entropy coding. To improve the coding gain, new features such as more intra modes and larger transform kernels have been developed. However, the coding components are optimized separately thus the joint optimization might lead to a higher compression ratio. In addition, the transform kernels such as discrete cosine transform (DCT) are performed linearly, thus adopting the non-linear feature is potential for improving the compression ability.

Recently, deep learning technology has demonstrated its powerful ability in many computer vision tasks such as \cite{krizhevsky2012imagenet, he2016deep, googlenet}. For the image compression, quite an amount of learning-based methods \cite{li2018fully, sun2020enhanced, dumas2019context, hu2019progressive, ma2019image, jia2019content, dai2017convolutional} have also been developed to improve the coding gain. About the end-to-end (E2E) learned image compression (LIC), RNN structures were utilized in \cite{toderici2017full} to perform a scalable coding. Convolutional autoencoder (CAE)-based methods have been greatly proposed in \cite{cheng2019learning, balle2018integer, Cheng2019EnergyCI, liu2019non, cheng2020learned}. For these CAE-based approaches, the difference between input images and reconstructed images is regarded as the distortion, and the size of the latent node represents the required bits. By the efficient network structure and enhanced entropy model, the E2E LIC has reached the same coding gain with BPG in terms of both MS-SSIM and PSNR metrics.

Along with a considerable coding gain improvement, the network complexity of LIC also increases hugely. First, to store and transfer a huge amount of parameters and activations, the memory consumption is high. Second, the arithmetic operation of weights and activations is computationally intensive when using 32-bit floating point. To alleviate the above two problems, pruning and quantization are two mainstream approaches. By using the pruning methods in \cite{han2015learning, li2016pruning, he2017channel, luo2017thinet}, some weights and activations are cut so that the memory consumption can be saved. However, the computational cost in the worst case still remains the same as origin which requires the 32-bit floating-point multiplier and adder. By using the quantization method, the bit-depth of weights and activations is reduced to save the memory consumption. Moreover, the computational burden can also be relieved. Though there have been tremendous pruning and quantization studies, the majority are for the well-known networks such as AlexNet \cite{krizhevsky2012imagenet} and ResNet \cite{he2016deep}, while there are very few works \cite{balle2018integer, nick2019computationally} for LIC. As reported in \cite{ma2019image}, the memory and computation efficient design for practical image and video codec is highly required.

In this paper, we propose an E2E LIC with each weight and activation element being quantized to a fixed-point arithmetic. The main contribution of this paper are as follows.

1) We first determine the optimal grouping and quantization codebook for the weight. After that, we propose a weight-clipping fine tuning scheme which can significantly improve the coding gain after the quantization.

2) To saturate the quantization range for the output activation, a coding gain-oriented non-linear quantization codebook is proposed which can increase the quantization precision for the certain range of the activations.

3) We propose a mean-reduced quantization scheme for the analysis transform outputs. By doing so, the bit-depth for the specific channel with large magnitude of non-zero mean can be reduced.

The experimental results show that when quantizing to 8-bit storage for both weight and activation, we can reach better coding gain than BPG in terms of MS-SSIM. About the hardware cost, we can save around 75\% memory consumption compared with the previous work.

\section{Related Works}

\subsection{Learned Image Compression}

\begin{figure*}[htb]
	\centering
	\includegraphics[height=8.5cm]{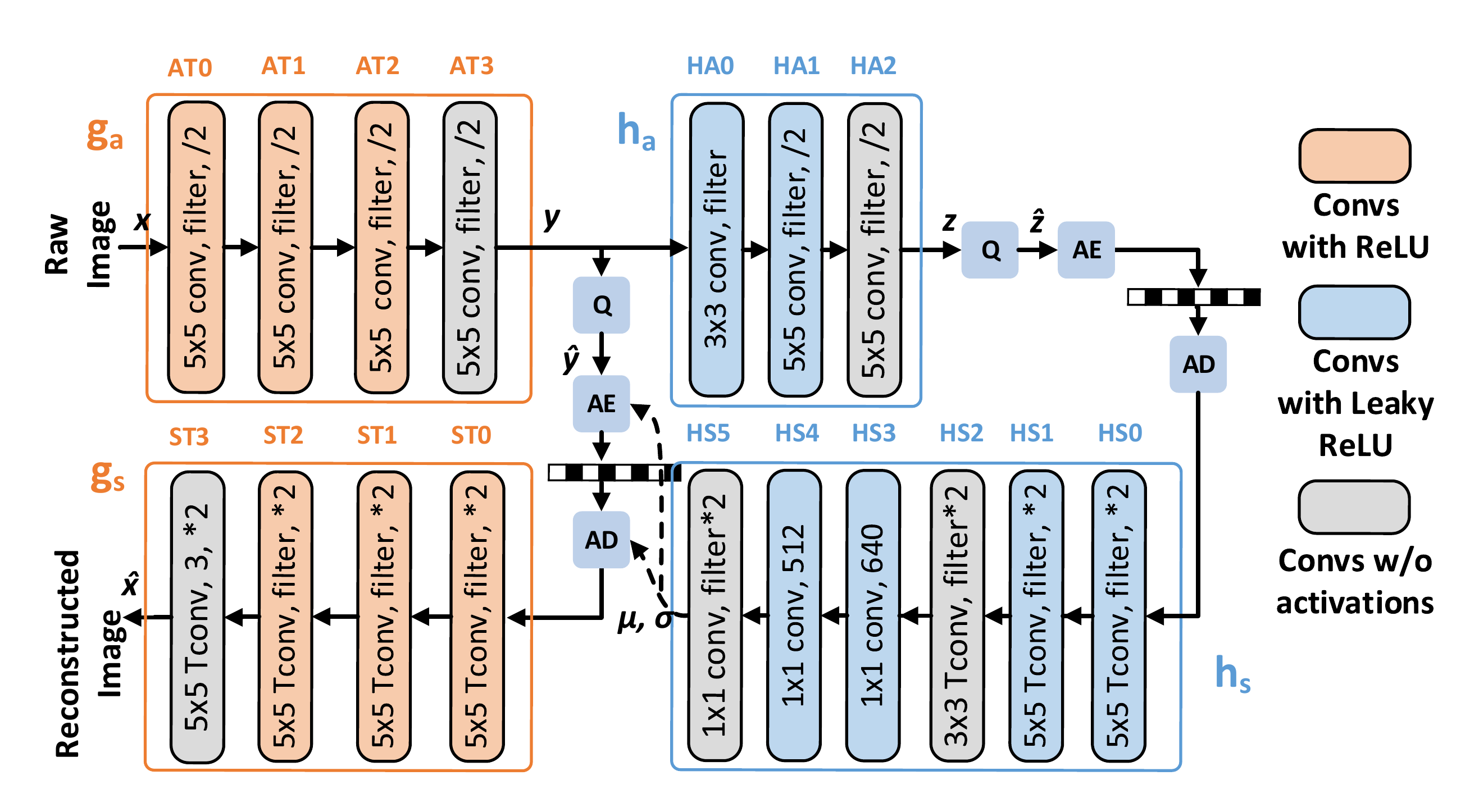}
	\caption{Diagram of mean-scale hyperprior baseline network, which is the \textit{hyperprior-5} in \cite{cheng2019deep}. The main path is composed of analysis transform $g_a$ (AT0-AT3) and synthesis transform $g_s$ (ST0-ST3). The hyper path is composed of analysis transform $h_a$ (HA0-HA2) and synthesis transform $h_s$ (HS0-HS5).}
	\label{fig_hyper}
\end{figure*}

CAE is the most popular framework for the LIC. Based on a well-designed network architecture, CAE with factorized prior \cite{balle2016end} can reach a comparable coding gain with JPEG2000. With a more powerful entropy model called \textit{hyperprior}, \cite{balle2018variational} can achieve the BPG performance. In this work, we also build the baseline network based on the \textit{hyperprior}, which is called \textit{hyperprior-5} in \cite{cheng2019deep} as shown in Fig. \ref{fig_hyper}. The operation can be formulated as the following two equations
\begin{equation}
\begin{split}
& y=g_a(x;\phi) \\
& \hat{y}=Q(y) \\
& \hat{x}=g_s(\hat{y};\theta) \\
\end{split}
\label{eq_main}
\end{equation}
\begin{equation}
\begin{split}
& z=h_a(y;\phi_h) \\
& \hat{z}=Q(z) \\
& \mu_y,\sigma_y=h_s(\hat{z};\theta_h)
\end{split}
\label{eq_hyper}
\end{equation}

\noindent where Eq. \ref{eq_main} and Eq. \ref{eq_hyper} represents the main and hyper path operation, respectively. For the main path, $x$ is the input image and $g_a$ is the analysis transform which transforms $x$ to the compressed code $y$. During the inference, $Q$ is the rounded operation to generate the quantized code $\hat{y}$ for the arithmetic coding. During the training, the rounding operation is replaced by adding a uniform noise $\mathcal{U}(-\tfrac{1}{2},\tfrac{1}{2})$. Finally, $\hat{y}$ is passed through the synthesis transform $g_s$ to generate the reconstructed image $\hat{x}$. For the hyper path, the operation flow is similar with the main path which contains the analysis transform $h_a$ and synthesis transform $h_s$, while the target of the synthesis transform is to generate the mean $\mu_y$ and scale $\sigma_y$ for the arithmetic coding of $\hat{y}$. In Eq. \ref{eq_main} and Eq. \ref{eq_hyper}, $\phi$, $\theta$, $\phi_h$ and $\theta_h$ are the network parameters which will be optimized during the training.

Same with traditional image coding, learned image compression adopts a Lagrangian multiplier-based rate-distortion optimization as shown in Eq. \ref{eq_rdcost}.
\begin{equation}
\begin{split}
\emph{J} & = R + \lambda \cdot D \\
& = \mathbb{E}[-log_2(p_{\hat{y}}(\hat{y}))] + \mathbb{E}[-log_2(p_{\hat{z}}(\hat{z}))]  + \lambda \cdot D(x,\hat{x}) \\
\end{split}
\label{eq_rdcost}
\end{equation}
where $\lambda$ is altered to balance the rate and distortion so that multiple models with various rates can be generated, $D(x,\hat{x})$ is the distortion between the original and reconstructed image which can be mean square error (MSE) or MS-SSIM \cite{wang2003multiscale}, $p_{\hat{y}}(\hat{y})$ is the probability distribution of $\hat{y}$ which is conditioned on a hyperprior as illustrated in \cite{balle2018variational} and \cite{minnen2018joint}. For each element $\hat{y_i}$, it is modeled as a Gaussian distribution $\mathcal{N}(\mu_{y_i}, \sigma_{y_i}^2)$ as shown in the following equation
\begin{equation}
p_{\hat{y}|\hat{z}, \theta_h}(\hat{y}|\hat{z},\theta_h) = \prod_{i} \big( \mathcal{N}(\mu_{y_i}, \sigma_{y_i}^2)*\mathcal{U}(-\frac{1}{2},\frac{1}{2}) \big)(\hat{y}_i)
\label{eq_y}
\end{equation}
\noindent where the element-wise mean $\mu_{y_i}$ and scale $\sigma_{y_i}$ are generated by the transform $h_s(\hat{z};\theta_h)$. For the quantized code $\hat{z}$, since there is no prior information for the probability distribution, a non-parametric factorized density model \cite{balle2018variational} is used as shown in the following equation where $\psi$ are the trainable parameters of each univariate distribution in the factorized density model.
\begin{equation}
p_{\hat{z}|\psi}(\hat{z}|\psi) = \prod_{i} \big( p_{z_i|\psi^{(i)}}(\psi^{(i)}) *\mathcal{U}(-\frac{1}{2},\frac{1}{2}) \big)(\hat{z}_i)
\label{eq_z}
\end{equation}

For the activation function, ReLU and Leaky-ReLU rather than the generalized divisive normalization (GDN) described in \cite{balle2016end} is used to ease the hardware implementation in the future.

\subsection{Neural Network Quantization}

Quantization is one of the most efficient ways to alleviate the memory and computational burden. After the quantization, floating-point precision values are usually represented by the fixed-point arithmetic or the integer with limited bit-depth.

About the weight quantization, Han \textit{et al.}\cite{han2015deep} clustered weights and then applied a non-uniform quantization. An incremental network quantization scheme was proposed in \cite{zhou2017incremental} and the key concept was to quantize the partial weights and then compensate the quantization loss by fine tuning the remaining weights recursively, until all the weights have been quantized. Vector quantization was utilized to compress the weights in \cite{gong2014compressing}. Lipschitz constraint-based width-level and multi-level network quantization were proposed for high-bit and low-bit quantization respectively in \cite{xu2019iterative}.

In addition, there are quite a few of studies aiming at the aggressive bit reduction. A binary weight network was proposed in \cite{courbariaux2015binaryconnect}, deterministic and stochastic binarization is exploited to generate a set of \{+1,-1\} weights. \cite{rastegari2016xnor} introduced a scaling factor for the binary weight, and decided the factor by minimizing the Euclidian distance between the original full-precision weights and scaled binary weights. Similar to \cite{rastegari2016xnor}, \cite{li2016ternary} quantized the weights to \{+1,0,-1\}. Another ternary weight network was presented in \cite{zhu2016trained}, and the threshold for the ternary partition and the ternary codebook could be obtained through the learning. A sparse ternary weight network was also proposed in \cite{jin2018sparse}. 

Different from merely quantizing the weight, many literature also studied on the activation quantization. Park \textit{et al.}\cite{park2018value} presented value-aware quantization so that lower quantization precision (QP) is applied to smaller values. The author also exploited the concept of weighted entropy and analyzed the impact of different weight/activation value on the final accuracy in \cite{park2017weighted}. Mishra \textit{et al.}\cite{mishra2017wrpn} proposed a low precision network providing more filters than the origin which can surpass the accuracy of the baseline full-precision network. There are also several works emphasizing on optimizing the activation range. Jung \textit{et al.} \cite{jung2019learning} determined the efficient activation interval by learning the center and distance parameter. Variants of ReLU with different bounding value and piece-wise function was proposed in \cite{cai2017deep}. Batch normalization \cite{ioffe2015batch} was utilized to purposely generate a Gaussian-like distribution and Lloyd's algorithm \cite{lloyd1982least} was used to optimize the quantization steps. Different with \cite{cai2017deep} whose clipping value is fixed, Choi \textit{et al.}\cite{choi2018pact} optimized the clipping range of activations through training and then linearly quantized both weights and activations to 4-bit. Zhou \textit{et al.} \cite{zhou2019progressive} equipped a low-precision network with a full-precision network and then gradually remove the impact of the full-precision network during the training.

Similar to the weight quantization, ultra-low bit precisions such as 1-bit and 2-bit were proposed for the activation. A binary network with 1-bit weight and activation was proposed in \cite{courbariaux2016binarized}. All the values are constrained to either +1 or -1 and the network is trained from scratch by using the straight-through estimator (STE) \cite{hinton2012neural}. Besides, \cite{rastegari2016xnor} also developed a XNOR-Net composed of binary weights and activations. To improve the performance of XNOR-Net, \cite{li2017performance} performed the binary quantization for the quantization residual iteratively. Mellempudi \textit{et al.}\cite{mellempudi2017ternary} ternarized the parameters by an improved theoretical formulation and reduced the bit-depth of activation to 4/8-bit.

Though quantization has achieved good performance across many popular network models, there are very few works for LIC. Ball{\'e} \textit{et al.} \cite{balle2018integer} designed an integer network and provided a very heuristic training scheme. The results show that exploiting an integer prior will not hurt the performance. However, using integer networks for both main and hyper path will diminish the coding gain which can be compensated by using more filters. Nick \textit{et al.} \cite{nick2019computationally} pruned the filters by subjoining the group Lasso regularization into the rate-distortion loss function. In addition, a low-complexity divisive normalization without the square and root operation was designed. Weight was quantized in \cite{sun2020end} and \cite{sun2021learned} while the activation in the main path was not quantized.

\section{Weight Quantization}

\subsection{Quantization with Dynamic Fixed-Point Representation}

\begin{figure*}[htb]
	\centering
	\includegraphics[height=2.8cm]{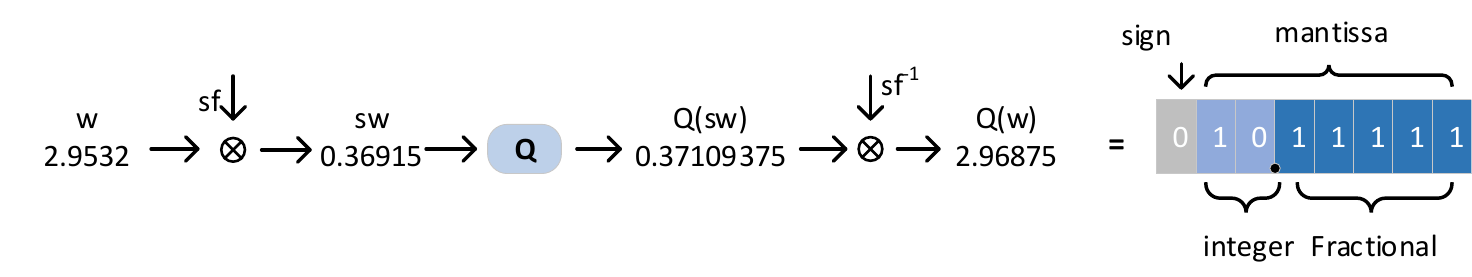}
	\caption{Quantization process for an arbitrary value such as weight. First, the weight $w$ is scaled to $sw$ by Eq. \ref{eq_scale}, and the scaling factor $\mathit{sf}$ can be calculated by Eq. \ref{eq_sf1}. After the scaling, $sw$ can be quantized to $Q(sw)$ as shown in Eq. \ref{eq_lq}. Finally, $Q(sw)$ is divided by $\mathit{sf}$ to obtain $Q(w)$ as described in Eq. \ref{eq_sf-1}, which can be represented by a dynamic 8-bit fixed-point format with sign, integer and fractional part.}
	\label{fig_fix_point}
\end{figure*}

The most common data format is the 32-bit floating point with IEEE 754 standard which has 1-bit for sign, 8-bit for exponent and 23-bit for mantissa. In our work, we aim at the 8-bit storage for each weight and activation element considering that 8-bit is friendly for the memory allocation in most hardwares. We adopt the dynamic fixed-point arithmetic in \cite{sze2017efficient}, and one example is shown in Fig. \ref{fig_fix_point}.

For an arbitrary value $w \in \mathbb{R}^N$, it is first scaled to $sw \in (-\alpha,\alpha)$ with a scaling factor ($\mathit{sf}$) as shown in Eq. \ref{eq_scale}
\begin{equation}
sw = w \times sf \\
\label{eq_scale}
\end{equation}
\noindent where $\alpha$ is set as 0.5 to make the dynamic range of $sw$ become one. To implement the scaling by shifting operation in the hardware, $\mathit{sf}$ is limited to the power of two, and can be calculated by Eq. \ref{eq_sf1}.
\begin{equation}
\mathit{sf} =  2 ^ { - \lfloor \log _2{ | w | ) } \rfloor - 1} \times \alpha 
\label{eq_sf1}
\end{equation}
The second step is to quantize $sw$ to $Q(sw)$. In the case of simply using linear quantization with N-bit, quantized value can be obtained by the following equation.
\begin{equation}
\small
Q(sw)=  \dfrac{  round(sw\times2^{N})   }{2^{N}} 
\label{eq_lq}
\end{equation}

Finally, $Q(w)$ can be obtained by dividing the scaling factor as shown in the below.
\begin{equation}
\small
Q(w)=  Q(sw)  \times sf^{-1}
\label{eq_sf-1}
\end{equation}

About the scaling, in the real implementation, scaling the weights in the element-wise will consume huge memory of $\mathit{sf}$. Therefore, weights are clustered to several groups thus each group can share one $\mathit{sf}$. For the group of weights $\boldsymbol{w}$, the $\mathit{sf}$ can be obtained by Eq. \ref{eq_sf2}. After the scaling, the group of scaled weights is denoted as $\boldsymbol{sw}$. Considering the structural feature of convolutional neural networks, channel-wise and layer-wise grouping are presented in the next section.
\begin{equation}
\mathit{sf} =  2 ^ { - \lfloor \log _2 { max ( | {\boldsymbol{w}} | ) } \rfloor - 1} \times \alpha 
\label{eq_sf2}
\end{equation}
About the quantization, there are linear quantization (LQ) and non-linear quantization (NLQ). The LQ example has been shown in Eq. \ref{eq_lq} where the distance between each two neighboring quantized values is a constant. When using NLQ, higher precision is provided for the interval of more importance. There are usually two ways for the NLQ, one is to utilize Lloyd's method \cite{lloyd1982least} to minimize the K-means of original and quantized values, which can be described in Eq. \ref{eq_kmeans}.
{
The objective of K-means clustering is to find
\begin{equation}
\small
\mathop{\arg\min}_{\bm{S}} \ \ \sum_{i=1}^{k}\sum_{x\in{S_i}}\| x-\mu_i \|^2
\label{eq_kmeans}
\end{equation}
\noindent where $k$ is the number of clustering, $\bm{S}=\{S_1,S_2,...,S_k\}$ are clustering sets, $\mu_i$ is the mean of points in $S_i$.	
}
The other method is to generate a piecewise LQ and the precision is different for each piece, as shown in Eq. \ref{eq_nlq}
\begin{equation}
\small
Q(sw_i)=  \dfrac{  round(  sw_i\times2^{N_i} )    }{2^{N_i}} , \text{if } |sw_i|\in[K_i,K_{i+1})
\label{eq_nlq}
\end{equation}
\noindent where $2^{-N_i}$ is the QP for the i-th interval. The length and QP of intervals should satisfy the following equation
\begin{equation}
\small
\sum_{i} (K_{i+1}-K_i) \times 2^{N_i} \times 2 = 2^N
\label{eq_quantize}
\end{equation}
\noindent where $[K_i,K_{i+1})$ is the interval and $2^{-N_i}$ is the QP for each piece, respectively, and $N$ is the bit budget for the storage of one element.

\subsection{Quantization and Grouping Scheme Determination}

\begin{figure}[t]
	\footnotesize
	\begin{minipage}[b]{1.0\linewidth}
		\centering
		\centerline{\includegraphics[width=9.5cm]{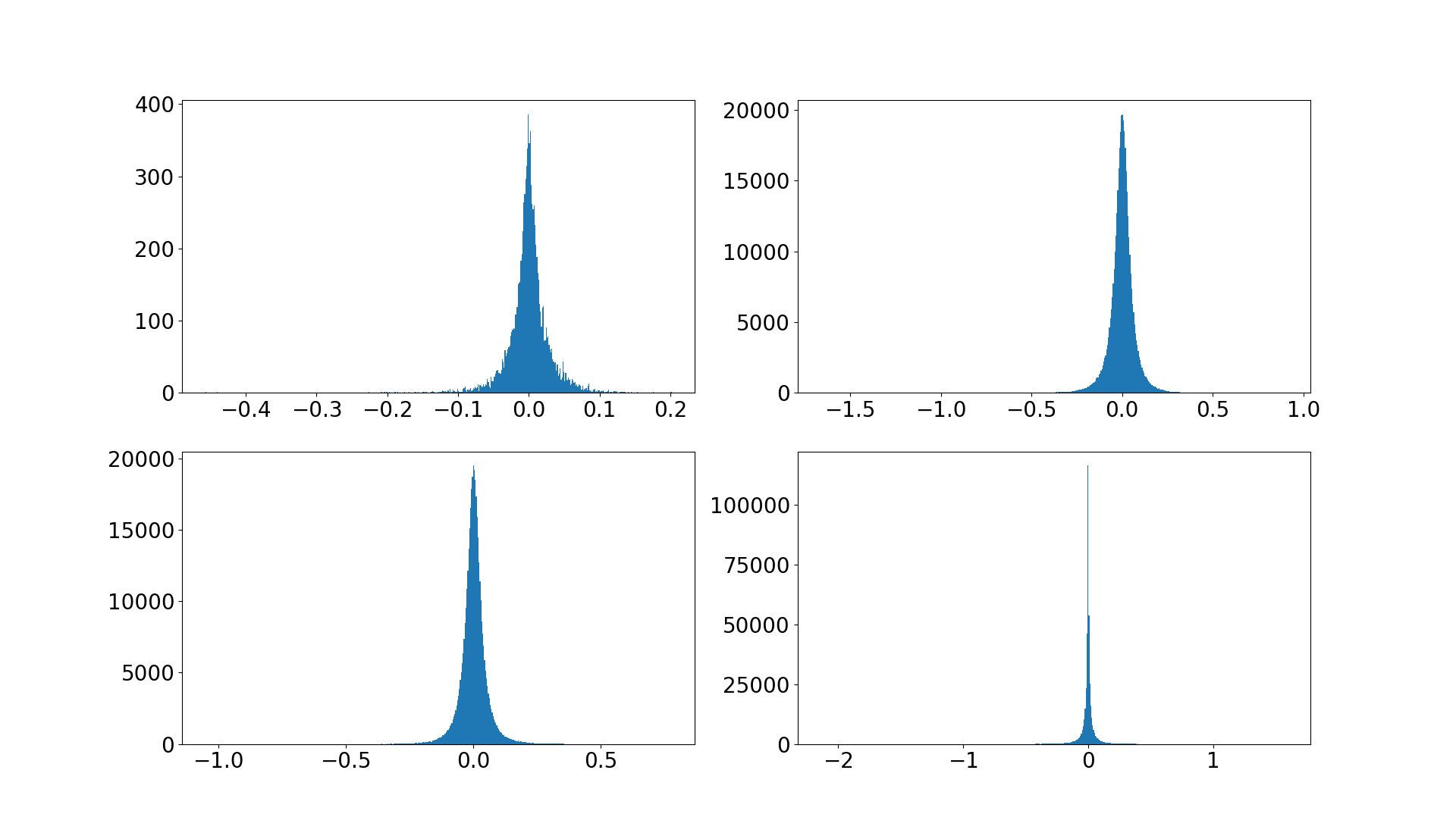}}
		\centerline{ (a) Four layers of analysis transform $g_a$.}
	\end{minipage}
	\begin{minipage}[b]{1.0\linewidth}
		\centering
		\centerline{\includegraphics[width=9.5cm]{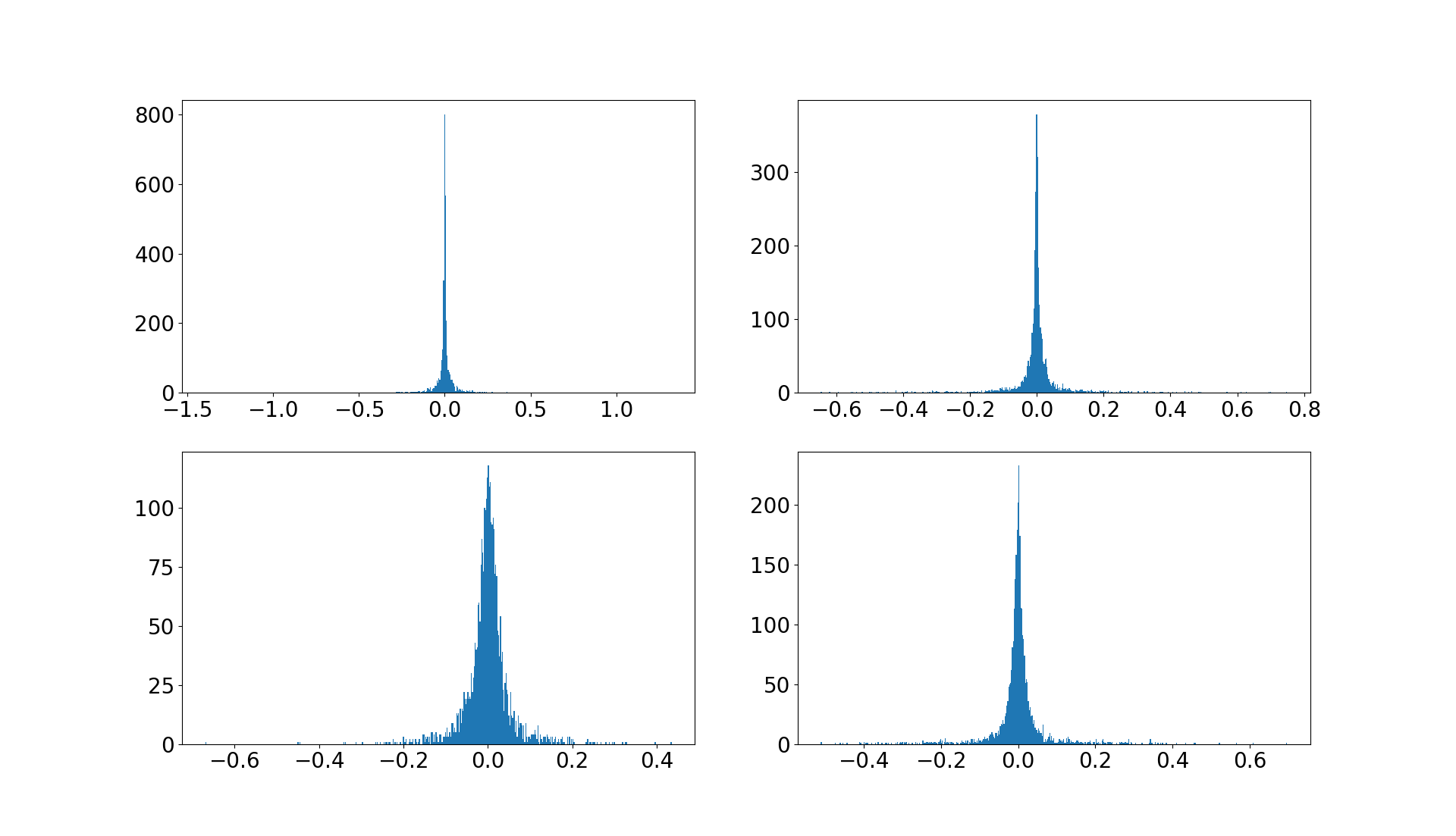}}
		\centerline{  (b) First four channels of the final layer of analysis transform $g_a$.}
	\end{minipage}
	\caption{Layer-wise and channel-wise distribution of weight value. Both follow Gaussian distribution.}
	\label{fig_weight_dist}
\end{figure}

To determine the quantization and grouping, we visualize the weight distribution at first. Four layers of analysis transform $g_a$ and the first four channels of the first layer of $g_a$ are shown in Fig. \ref{fig_weight_dist}. We can see that the distribution follows the {Laplace} distribution well so that higher precision can be offered for the values closer to zero. In our method, we develop a simple NLQ codebook as shown in the following equation
\begin{equation}
\small
Q(sw) = \begin{cases}
 \dfrac{ round( sw\times2^{N-1} )  }{2^{N-1}}   , & \text{if } |sw|\in[ \frac{\alpha}{4} ,\alpha)\\
 \dfrac{ round( sw\times2^{N} )  }{2^{N}}  , & \text{if } |sw|\in[\frac{\alpha}{8}, \frac{\alpha}{4})\\
 \dfrac{ round( sw\times2^{N+2} )  }{2^{N+2}} , & \text{if } |sw|\in[0, \frac{\alpha}{8})\\
\end{cases}
\label{eq_nlq2}
\end{equation}
\noindent where $N$ is the bit budget and $\alpha$ is the dynamic range.

As described in the previous sub-section, there are two kinds of scaling: channel-wise (CW) and layer-wise (LW). In the case of LW, each group contains the weight $\boldsymbol{w} \in R^{I \times W \times H \times O}$ where four dimension represents input channel number, kernel width, kernel height and output channel number. In the case of CW, each group contains the weight $\boldsymbol{w} \in R^{I \times W \times H}$. For both grouping scheme, we attempt the LQ in Eq. \ref{eq_lq} and NLQ in Eq. \ref{eq_nlq}. Besides, we also exploit Eq. \ref{eq_lloyd} for LW rather than CW since it is not feasible to memorize LUTs for all the channels. At a moderate rate, the average coding gain results of Kodak images \cite{franzen1999kodak} are shown in Table \ref{table_1}. From the results, we can conclude that using CW is better than LW. It is because different channels have various dynamic ranges as shown in Fig. \ref{fig_weight_dist}, so using LW will quantize more weights to zero. About the quantization scheme, NLQ is apparently better than LQ in terms of the coding gain, which is consistent with the value distribution.

\begin{table}[t]
	\setlength{\abovecaptionskip}{0.0cm}   
	\setlength{\belowcaptionskip}{0.0cm}
	\caption{Comparison of different quantization and grouping scheme in terms of coding gain}
	\footnotesize 
	\begin{center}
		{
			
			\resizebox{0.95\width}{!}{
			
			\begin{tabular}{|c|c|c|c|c|c|c|}
				\hline
				
				\textbf{ }					&\text{ }			&\multicolumn{3}{|c|}{\text{Layer-wise (LW)}}			   		 &\multicolumn{2}{|c|}{\text{Channel-wise (CW)}}				\\ \hline
				\textbf{ }					&\text{Origin}		&\text{LQ}			&\text{NLQ}			&\text{Lloyd}		&\text{LQ} 			&\textbf{NLQ}			\\ \hline
				
				\textbf{PSNR (dB)}			&\text{$32.32$}		&\text{$28.81$}		&\text{$32.08$}		&\text{$32.18$}		&\text{$31.60$} 		&\boldmath{$32.25$}		\\ \hline
				
				\textbf{bpp}				&\text{$0.529$}		&\text{$0.770$}		&\text{$0.555$}		&\text{$0.537$}			&\text{$0.569$} 		&\boldmath{$0.539$}		\\ \hline
				
			\end{tabular}
		
			}
		
		}
	\end{center}
	\label{table_1}
	\vspace{-4mm}
\end{table}

\begin{table}[t]
	\setlength{\abovecaptionskip}{-0.5cm}   
	\setlength{\belowcaptionskip}{0.0cm}
	\caption{Comparison of different highest quantization precision in terms of coding gain (N=8)}
	\footnotesize
	\begin{center}
		{
			\begin{tabular}{|c|c|c|c|c|c|c|}
				\hline
				\textbf{Precision}					&\text{$2^{-N}$}		&\text{$2^{-(N+1)}$}	&\text{$2^{-(N+2)}$}		&\text{$2^{-(N+3)}$}			\\ \hline
				
				\textbf{PSNR (dB)}			&\text{$31.60$}		&\text{$32.17$}		&\text{$32.28$}		&\text{$32.30$}		\\ \hline
				
				\textbf{bpp}				&\text{$0.569$}		&\text{$0.541$}		&\text{$0.532$}		&\text{$0.530$}			\\ \hline
				
			\end{tabular}
		}
	\end{center}
	\label{table_2}
	\vspace{-4mm}
\end{table}

Through the above preliminary experiments, we have found that CW-NLQ with the highest precision $2^{-(N+2)}$ can achieve the best coding gain. To ensure that this highest precision is enough, we evaluate the CW-LQ with different precisions as shown in Table \ref{table_2}. We can see that when increasing the precision from $2^{-N}$ to $2^{-(N+2)}$, PSNR will be improved and rate can be reduced. However, when further using one more bit, there is only 0.02dB and 0.002bpp improvement, which is quite trivial.

\subsection{Weight Clipping Fine Tuning (WCFT) Scheme}

By using proposed CW-NLQ, the coding loss is not negligible especially for high rate models. To solve this problem, we propose a weight clipping fine tuning (WCFT) method.

For each group of weights $\boldsymbol{w}$, the quantization error can be formulated as Eq. \ref{eq_error}
\begin{equation}
\small
\text{Quantized error} = \sum_{i=0}^{ card(\boldsymbol{sw})-1} ||sw_i-Q(sw_i)||_2^2 \times sf^{-1}
\label{eq_error}
\end{equation}
\noindent where $sw_i$ represents each scaled weight element. Given a fixed quantization codebook, $|sw_i-Q(sw_i)|$ is fixed. Therefore, to reduce the quantization error, a large $\mathit{sf}$ is demanded. To enlarge the value of $\mathit{sf}$, $max|\boldsymbol{w}|$ should be reduced as shown in Eq. \ref{eq_sf2}. Denote the weights after the clipping is $\boldsymbol{w}'$ and the corresponding scaling factor is $\mathit{sf'}$, we have the following relationship
\begin{equation}
\small
\begin{aligned}
sf &\rightarrow sf' \triangleq sf \times 2 \\
max(|\boldsymbol{w}|) &\rightarrow max(|\boldsymbol{w'}|)  \triangleq  2 ^ { \lfloor \log _2  { max ( |\boldsymbol{w}| ) } \rfloor} - \epsilon\\
\end{aligned}
\label{eq_nlq3}
\end{equation}
\noindent where $\epsilon$ is a very small positive value which can be set by the following analysis. First, $\mathit{sf'}$ and $max(|\boldsymbol{w'}|)$ observes the following equation.
\begin{equation}
\small
\begin{split}
sf' \times max(|\boldsymbol{w'}|) =   max(| {  \boldsymbol{sw'} } |) \\
\end{split}
\label{eq_nlq4}
\end{equation}

When using the proposed NLQ in Eq. \ref{eq_nlq2}, the maximum $|\boldsymbol{sw'}|$ that can be remained after the quantization is $\alpha - 2^{-(N-1)}$. Therefore, we have the following equation.
\begin{equation}
\small
\begin{split}
max(| {  \boldsymbol{sw'} } |) \triangleq \alpha - 2^{-(N-1)} 
\end{split}
\label{eq_nlq5}
\end{equation}

Substituting Eq. \ref{eq_nlq3} into Eq. \ref{eq_nlq4}, we can obtain the value of $\epsilon$ and the clipping function.
\begin{equation}
\small
\begin{split}
\epsilon = 2 ^ { \lfloor \log _2  { max ( | {\boldsymbol{w}} | ) } \rfloor - (N-1) } \times \alpha^{-1}
\end{split}
\label{eq_epsilon}
\end{equation}

\begin{equation}
\small
f_c(w)=\begin{cases}
w, &\text{  if } w \leqslant 2 ^ { \lfloor \log _2  { max ( | {\boldsymbol{w}} | ) } \rfloor} - \epsilon\\
2 ^ { \lfloor \log _2  { max ( | {\boldsymbol{w}} | ) } \rfloor} - \epsilon, &\text{  otherwise}
\end{cases}
\label{eq_clip}
\end{equation}

To fine-tune the weights gradually, the clipping can be executed for several rounds with a factor $\beta$ ($\beta \geqslant 1$) as shown in Eq. \ref{eq_clip_beta}.
\begin{equation}
\small
f_c(w, \beta)=\begin{cases}
w, \text{  if } w \leqslant 2 ^ { \lfloor \log _2  { max ( | {\boldsymbol{w}} | ) } \rfloor} \times \beta - \epsilon\\
2 ^ { \lfloor \log _2  { max ( | {\boldsymbol{w}} | ) } \rfloor} \times \beta - \epsilon, \text{  otherwise}
\end{cases}
\label{eq_clip_beta}
\end{equation}

The proposed training scheme is shown in \textbf{Algorithm 1}. First, we train the network for $I_1$ iterations to obtain the optimized weights as usual. The loss function is shown in Eq. \ref{eq_rdcost}. After obtaining the pre-trained weights, we group the weights to channel-wise $\boldsymbol{w} \in R^{I \times W \times H}$. For each element $w$, we clip the value to $f_c(w, \beta)$. After the clipping, we will fine tune the weights for $I_3$ iterations to compensate the clipping loss. The clipping and the subsequent fine tuning are conducted for $I_2$ rounds. The setting for $\{I_1, I_2, I_3\}$ are shown in the experimental results. During the backward propagation, we can obtain $\tfrac{\delta \emph{J}}{\delta f_c(w)}$ first and then calculate $\tfrac{\delta \emph{J}}{\delta w}$ by using straight-through estimator \cite{hinton2012neural}, that is to preserve the gradient and cancels the gradient when $w$ is larger than the threshold as shown in Eq. \ref{eq_ste}. Finally, we quantize $w$ to $Q(w)$ which will be used in the inference.

\begin{algorithm} [t]
	\small
	\caption{Proposed Training Method with Weight Clipping Fine Tuning}
	\label{alg:Framwork}
	\begin{algorithmic}[1]
		
		\For{ normal training iterations \textit{$I_1$} }
		
		\State Run forward propagation and compute \emph{J} in Eq. \ref{eq_rdcost}
		\State Run backward propagation and update $\boldsymbol{w}$ by descending SGD
		
		\EndFor
		
		\For { clipping rounds \textit{$I_2$} }
		
		\State Group weights to channel-wise $\boldsymbol{w} \in R^{I \times W \times H}$
		\State For each element $w$, clip $w$ to $fc(w, \beta)$ as shown in Eq. \ref{eq_clip_beta}
		
		\For{ fine tuning iterations \textit{$I_3$} }
		
		\State Run forward propagation and compute \emph{J} in Eq. \ref{eq_rdcost}
		\State Run backward propagation and obtain $\tfrac{\delta \emph{J}}{\delta f_c(w)}$
		\State Obtain $\tfrac{\delta \emph{J}}{\delta w}$ by Eq. \ref{eq_ste} and update $w$ by descending SGD
		\EndFor
		
		\EndFor
		
		\State Finally quantize $w$ to $Q(w)$
		\State \textbf{return} $Q(w)$
		
	\end{algorithmic}
\end{algorithm}
\begin{equation}
\small
\tfrac{\delta \emph{J}}{\delta w}=\begin{cases}
\tfrac{\delta \emph{J}}{\delta f_c(w)}, &\text{  if } w \leqslant 2 ^ { \lfloor \log _2 { max ( | {\boldsymbol{w}} | ) } \rfloor} \times \beta - \epsilon\\
0, &\text{  otherwise}
\end{cases}
\label{eq_ste}
\end{equation}

\section{Activation Quantization}

\subsection{Coding Gain-Oriented Non-linear Quantization Codebook}

The whole network is shown in Fig. \ref{fig_hyper} which is composed of main path and hyper path. There are four layers in the analysis transform $g_a$ and the synthesis transform $g_s$, respectively. For the hyper path, there are three layers for the analysis transform $h_a$ and six layers for the synthesis transform $h_s$. Our target is to quantize the activations (outputs) of these layers.

\begin{figure*}[t]
	\footnotesize
	\begin{minipage}[b]{1.0\linewidth}
		\centering
		\centerline{\includegraphics[height=2.7cm]{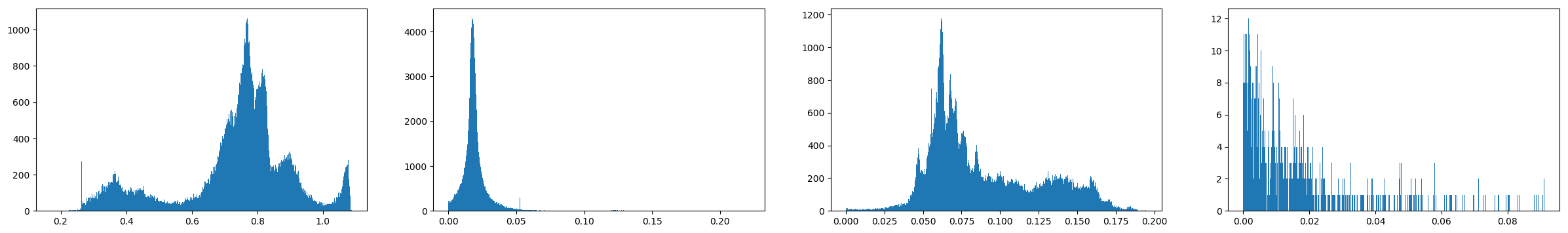}}
		\centerline{(a) Test image Kodak12.}
	\end{minipage}
	\begin{minipage}[b]{1.0\linewidth}
		\centering
		\centerline{\includegraphics[height=2.7cm]{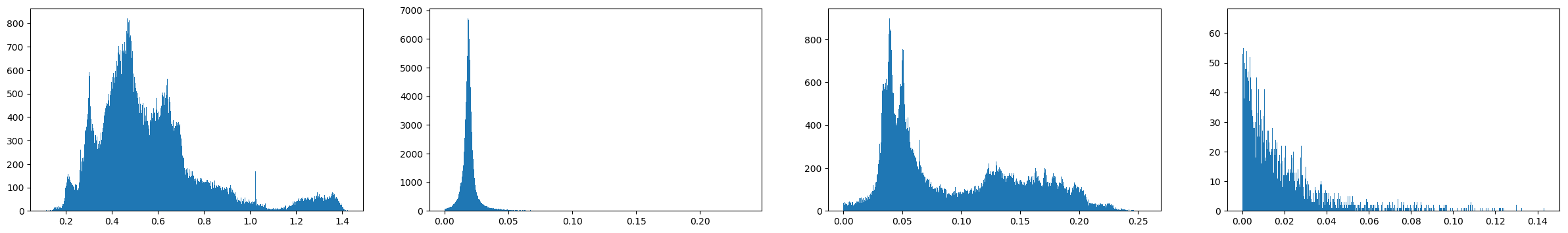}}
		\centerline{(b) Test image Kodak23.}
	\end{minipage}
	\caption{Activation distribution of the four channels of the shallowest layer (AT0) of the analysis transform. There is no unified distribution for various channels, and dynamic range also varies for different channels.}
	\label{fig_at0}
\end{figure*}

The distribution of four channels for two test images $Kodak12$ and $Kodak23$ are shown in Fig. \ref{fig_at0}. First, different from weight, the distribution of activations do not follow any clear distribution especially for the shallow layers. We can see that some channels follow Half-Gaussian or Gaussian, while others have no obvious shape. So we start from utilizing the LQ since it is hard to define a unified form for all the channels. Second, the dynamic range of various channels are different, thus CW rather than LW is adopted for the grouping scheme. In fact, this phenomenon is similar with DCT. In the case of using DCT, the dynamic range of different frequency components are various. DC component usually occupies the largest magnitude, while the AC components with high frequency are close to zero. Similar with the methods for DCT output truncation, we use several training images to determine the dynamic range of each channel.

For the activation $\boldsymbol{O} \in \mathbb{R}^{o_w \times o_h \times o_c}$ where three dimensions stand for the output feature map width, output feature map height and channel number, the scaling factor $\mathit{sf}$ for each channel can be calculated as 
\begin{equation}
sf =  2 ^ { - \lfloor \log _2  { max ( | {\boldsymbol{O}} | ) } \rfloor - 1} \times \alpha , \boldsymbol{O} \in \mathbb{R}^{o_w \times o_h}
\label{eq_sf_o}
\end{equation}

To set the dynamic range of scaled output ($so$) as one, $\alpha$ is 0.5 in the case of using leaky-ReLU or no activation, and it is 1.0 when using ReLU since only non-negative values will remain after ReLU. The scaling is executed as shown in Eq. \ref{eq_sf2}. However, when using CW-LQ, the coding gain loss is not ignorable especially for the large $\lambda$. To reduce the coding loss, we propose a coding gain-oriented NLQ codebook.

After the scaling, the maximum possible magnitude of $|so|$ is $\alpha$. However, $|so|$ rarely saturate to the largest dynamic range, thus some quantized steps will be wasted. In the case of a moderate bitrate model, we separate largest $|so|$ into four intervals $[\tfrac{\alpha}{2}, \tfrac{5\times \alpha}{8})$, $[\tfrac{5\times \alpha}{8}, \tfrac{3\times \alpha}{4})$, $[\tfrac{3\times \alpha}{4}, \tfrac{7\times \alpha}{8})$ and $[\tfrac{7\times \alpha}{8}, \alpha)$. The possibility of four intervals for the first two layers in AT are shown in Table \ref{table_4interval}. We can see that there is less than 20\% values saturate to the maximum magnitude $\alpha$.

\begin{table}[t]
	\setlength{\abovecaptionskip}{-0.5cm}   
	\setlength{\belowcaptionskip}{0.0cm}
	\caption{Probability of maximum $|so|$ located in different intervals for the first two layers in $AT$}
	\footnotesize
	\begin{center}
		{
			\begin{tabular}{|c|c|c|c|c|c|c|}
				\hline
				\textbf{Interval}					
				&$[\tfrac{\alpha}{2}, \tfrac{5\times \alpha}{8})$		
				&$[\tfrac{5\times \alpha}{8}, \tfrac{3\times \alpha}{4})$	
				&$[\tfrac{3\times \alpha}{4}, \tfrac{7\times \alpha}{8})$		
				&$[\tfrac{7\times \alpha}{8}, \alpha)$			\\ \hline
				
				\textbf{AT0}			
				&\text{$30.47 \%$}		
				&\text{$31.25 \%$}		
				&\text{$22.66 \%$}		
				&\text{$15.63 \%$}		\\ \hline
				
				\textbf{AT1}			
				&\text{$32.03\%$}		
				&\text{$27.34\%$}		
				&\text{$25.00\%$}		
				&\text{$15.63\%$}		\\ \hline
				
			\end{tabular}
		}
	\end{center}
	\label{table_4interval}
	\vspace{-4mm}
\end{table}

\begin{table*}[t]
	\setlength{\abovecaptionskip}{-0.5cm}   
	\setlength{\belowcaptionskip}{0.0cm}
	\caption{Coding gain improvement when increasing the QP from $\frac{1}{2^N}$ to $\frac{1}{2^{N+1}}$ for each interval}
	\footnotesize
	\begin{center}
		{
			\begin{tabular}{|c|c|c|c|c|c|c|c|c|c|c|}
				\hline
				\textbf{Interval}	
				&$[0, \tfrac{\alpha}{8})$				
				&$[\tfrac{\alpha}{8}, \tfrac{\alpha}{4})$		
				&$[\tfrac{\alpha}{4}, \tfrac{3\times \alpha}{8})$	
				&$[\tfrac{3\times \alpha}{8}, \tfrac{\alpha}{2})$		
				&$[\tfrac{\alpha}{2}, \tfrac{5\times \alpha}{8})$
				&$[\tfrac{5\times \alpha}{8}, \tfrac{3\times \alpha}{4})$		
				&$[\tfrac{3\times \alpha}{4}, \tfrac{7\times \alpha}{8})$		
				&$[\tfrac{7\times \alpha}{8}, \alpha)$		\\ \hline
				
				\textbf{Coding gain (dB)}			
				&\text{$0.080$}		
				&\text{$0.097$}		
				&\text{$0.074$}		
				&\text{$0.058$}	
				&\text{$0.036$}		
				&\text{$0.011$}		
				&\text{$0.002$}		
				&\text{$0.000$}	\\ \hline
				
			\end{tabular}
		}
	\end{center}
	\label{table_8interval}
	\vspace{-4mm}
\end{table*}

For the weight quantization, in order to saturate $sw$ to the maximum dynamic range, WCFT is proposed to clip $w$ to a certain range. However, activation results possess the physical information, thus it cannot be directly clipped. Taking DCT as an instance, clipping the dynamic range of DC component will hurt the coding gain drastically. Therefore, we utilize four possible dynamic ranges described above. For each channel, we use a multiplexer to select one possible dynamic range to fully utilize all the quantized steps.

When the maximum $|so|$ is located in $[\tfrac{7\times \alpha}{8}, \alpha)$, the dynamic range is equal to one thus a simple uniform quantization is adopted. In the case of N-bit budget, the QP is $\tfrac{1}{2^N}$ as shown in the following equation. 
\begin{equation}
\small
\text{Given } max(|so|) \in [\tfrac{7\times \alpha}{8}, \alpha),  \quad Q(so)=  \dfrac{   round(so\times2^{N} )     }{2^{N}} 
\label{eq_lq_o}
\end{equation}

When the maximum $|so|$ is located in $[\tfrac{3\times \alpha}{4}, \tfrac{7\times \alpha}{8})$, the quantization steps for $[\tfrac{7\times \alpha}{8}, \alpha)$ are not used any more. Therefore, we can increase the QP of a certain interval in  $[0, \tfrac{7\times \alpha}{8})$. The interval can be selected by the importance to the coding gain. We analyze eight intervals with the same range $\frac{\alpha}{8}$. For each interval $[\tfrac{i\times \alpha}{8}, \tfrac{(i+1) \times \alpha}{8})$, we analyze the coding gain improvement when increasing the QP from $\frac{1}{2^N}$ to $\frac{1}{2^{N+1}}$ and the results are shown in Table \ref{table_8interval}. We can see that the coding gain increasement coming from $[\tfrac{\alpha}{8}, \tfrac{\alpha}{4})$ is the largest. Therefore, the QP for this interval is increased for one more bit as shown in the following equation.

\begin{equation}
\small
\begin{split}
&\text{Given } max(|so|) \in [\tfrac{3\times \alpha}{4}, \tfrac{7\times \alpha}{8}) \\
&Q(so) = \begin{cases}
\dfrac{ round(so\times2^{N}) }{2^{N}}   ,  &\text{if } |so|\in[ \frac{\alpha}{4} ,\tfrac{7\times \alpha}{8})\\
\dfrac{ round(so\times2^{N+1}) }{2^{N+1}}  ,  &\text{if } |so|\in[\frac{\alpha}{8}, \frac{\alpha}{4})\\
\dfrac{ round(so\times2^{N}) }{2^{N}}  ,  &\text{if } |so|\in[0, \frac{\alpha}{8})\\
\end{cases} \\
\end{split}
\label{eq_nlq_175}
\end{equation}

Similarly, when the maximum $|so|$ is located in $[\tfrac{5\times \alpha}{8}, \tfrac{3\times \alpha}{4})$, the QP for the two intervals with the highest two coding gain improvements is increased by one bit. Therefore, the NLQ codebook is shown in Eq. \ref{eq_nlq_150}.

\begin{equation}
\small
\begin{split}
&\text{Given } max(|so|) \in [\tfrac{5\times \alpha}{8}, \tfrac{3\times \alpha}{4}) \\
&Q(so) = \begin{cases}
 \dfrac{ round(so\times2^{N}) }{2^{N}}  , & \text{if } |so|\in[\frac{\alpha}{4}, \tfrac{3\times \alpha}{4})\\
 \dfrac{ round(so\times2^{N+1}) }{2^{N+1}}  , & \text{if } |so|\in[0, \frac{\alpha}{4})\\
\end{cases}
\end{split}
\label{eq_nlq_150}
\end{equation}

When the maximum $|so|$ is located in $[\tfrac{\alpha}{2}, \tfrac{5\times \alpha}{8})$, the QP for the three intervals with the highest three coding gain improvements is increased by one bit. Based on the results in Table \ref{table_8interval}, three intervals are $[0, \tfrac{\alpha}{8})$, $[\tfrac{\alpha}{8}, \tfrac{\alpha}{4})$ and $[\tfrac{\alpha}{4}, \tfrac{3\times \alpha}{8})$.
\begin{equation}
\small
\begin{split}
&\text{Given } max(|so|) \in [\tfrac{\alpha}{2}, \tfrac{5\times \alpha}{8}) \\
&Q(so) = \begin{cases}
 \dfrac{ round(so\times2^{N}) }{2^{N}}  , & \text{if } |so|\in[\tfrac{3\times \alpha}{8}, \tfrac{5\times \alpha}{8})\\
 \dfrac{ round(so\times2^{N+1} ) }{2^{N+1}}  , & \text{if } |so|\in[0, \tfrac{3\times \alpha}{8})\\
\end{cases}
\end{split}
\label{eq_nlq_125}
\end{equation}

Noted that for the final layer of AT, $y$ is rounded to $\hat{y}$ in the element-wise in the original framework. We also follow this behavior. For the hyper path, $z$ is also rounded to $\hat{z}$ in the element-wise. For the activations in the hyper path, LQ is utilized since the bit reduction by using NLQ is quite trivial.

{
We use nine training images to determine the dynamic range of each channel. For the activation value out of the range, it is clipped. Noted that there are no overlap between these nine images and Kodak test set. From the experimental results, we can see that the coding performance loss is quite small when quantizing the activations, which shows the rationality of the determined dynamic range.
}

\subsection{Mean-Reduced Quantization for Analysis Transform Output}

\begin{figure*}[t]
	\footnotesize
	\begin{minipage}[b]{0.5\linewidth}
		\centering
		\centerline{\includegraphics[height=6.5cm]{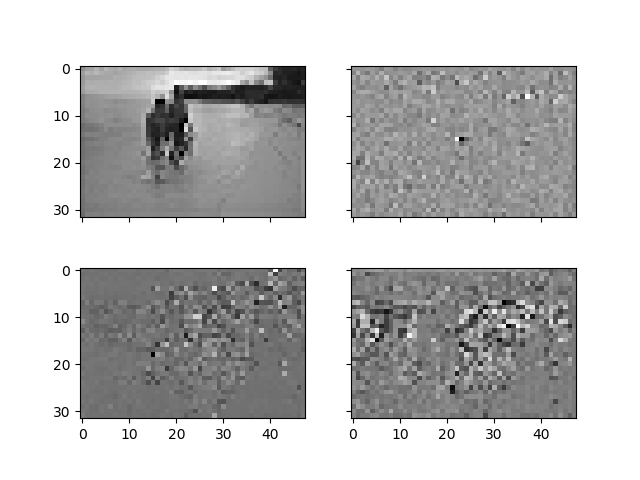}}
		\centerline{(a) Visualized $y$ for channel 135-138 of test image Kodak12.}
	\end{minipage}
	\begin{minipage}[b]{0.5\linewidth}
		\centering
		\centerline{\includegraphics[height=6.5cm]{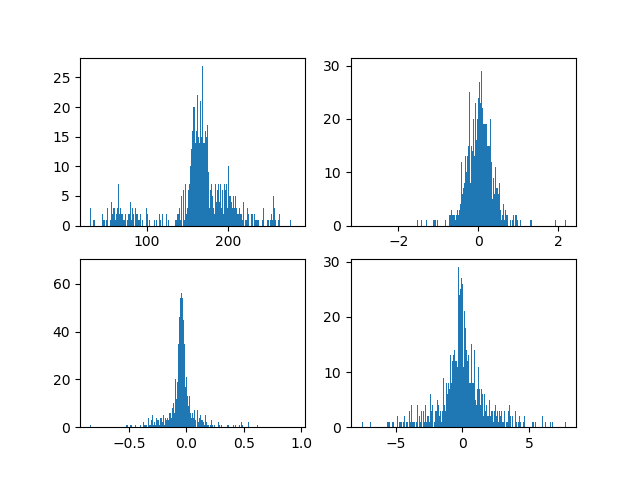}}
		\centerline{(b) Distribution for $y$ for channel 135-138 of test image Kodak12.}
	\end{minipage}
	\begin{minipage}[b]{0.5\linewidth}
	\centering
	\centerline{\includegraphics[height=6.5cm]{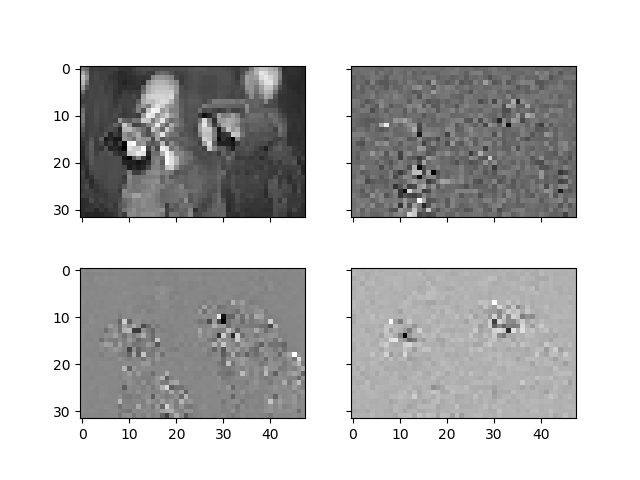}}
	\centerline{(c) Visualized $y$ for channel 135-138 of test image Kodak23.}
	\end{minipage}
	\begin{minipage}[b]{0.5\linewidth}
		\centering
		\centerline{\includegraphics[height=6.5cm]{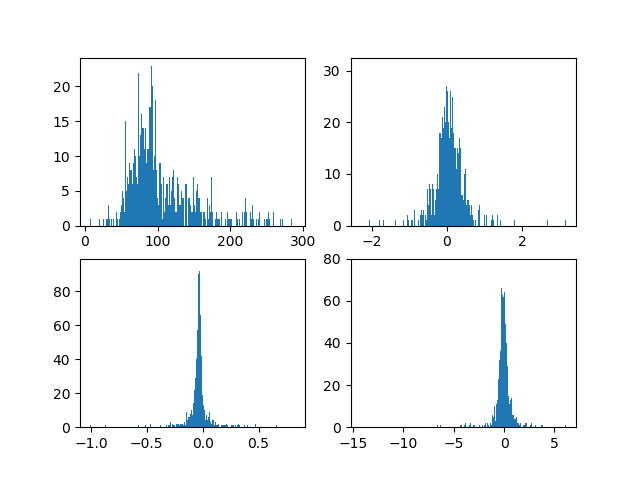}}
		\centerline{(d) Distribution for $y$ for channel 135-138 of test image Kodak23.}
	\end{minipage}
	\caption{Visualized results and value distribution of analysis transform output $y$ for the channel 135-138. In each figure (a)-(d), the left-top one represents the 135-th channel which owns the largest energy and non-zero mean.}
	\label{fig_at3}
\end{figure*}

\begin{figure*}[t]
	\centering
	\includegraphics[height=9.5cm, width=\linewidth]{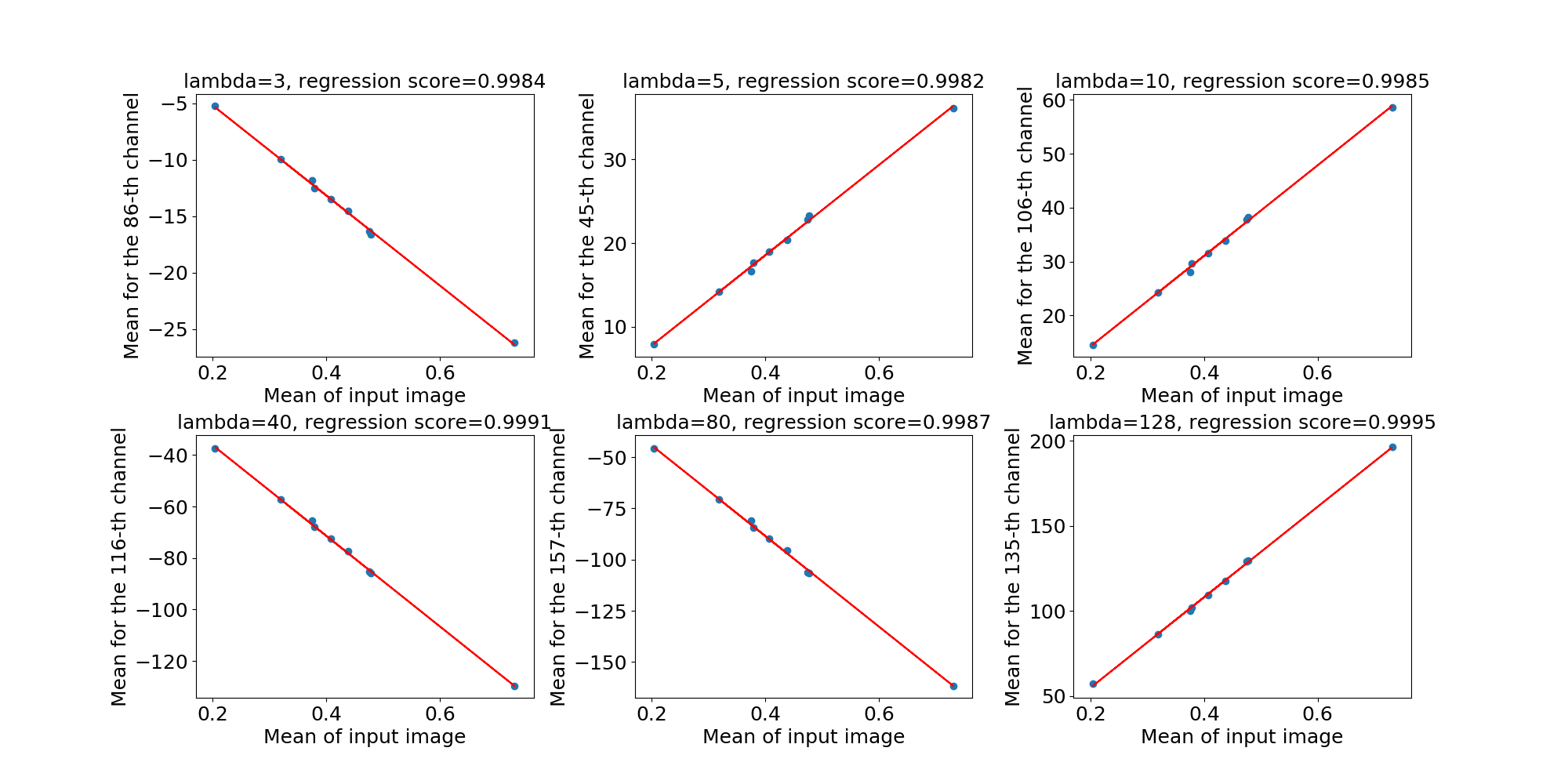}
	\caption{A linear regression between the mean of input image and the mean of the Gaussian distribution for the specific channel in the case of different $\lambda$.}
	\label{fig_mean}
\end{figure*}

In the LIC framework, AT and ST resembles the DCT and inverse DCT in the traditional image coding. After adopting the 2D DCT, the left-top component represents the DC component that is the mean of the input image. Inspired by this phenomenon, we first explore the channel-wise distribution of the AT results as shown in Fig. \ref{fig_at3}. We can observe that each channel is normally distributed which can be formulated as
\begin{equation}
\small
p(\boldsymbol{\hat{y}_k}) =  \mathcal{N}(\mu, \sigma^2)
\label{eq_y_hat}
\end{equation}
\noindent where $\boldsymbol{\hat{y}_k}$ is the values for the k-th channel, $\mu$ and $\sigma$ are the mean and scale. About the mean, for the 135-th channel, the magnitude is quite larger than the other channels. Moreover, the mean of this channel is non-zero while the mean of other sampled channels are quite close to zero. In the case of non-zero $\mu$, we have the following relationship.
\begin{equation}
\small
|max(\boldsymbol{\hat{y}_k}-\mu)| \leqslant |max(\boldsymbol{\hat{y}_k})|
\label{eq_y_hat_mean}
\end{equation}
Therefore, for each element $\hat{y_i}$ of the 135-th channel, $\hat{y_i}-\mu$ rather than $\hat{y_i}$ is stored in the memory. However, the problem is that when dealing with the AT results, the hyper path has not been decoded yet thus the mean has not been generated. Inspired by DCT, we find that $\mu$ has a linear relationship with the mean of the input image as formulated in Eq. \ref{eq_linear}
\begin{equation}
\small
\mu = a \times \bar{\boldsymbol{x}} + b
\label{eq_linear}
\end{equation}
\noindent where $\bar{\boldsymbol{x}}$ represents the mean of input image ${\boldsymbol{x}}$. The linear regression results of six $\lambda$ are shown in Fig. \ref{fig_mean}. We can see that the regression score is quite close to 1 for all the cases.

By storing $\hat{y_i}-\mu$ rather than $\hat{y_i}$ for the non-zero mean channel, the number of required bits per element can be reduced. For two higher $\lambda$ when optimizing MSE, the largest bit-depth for each element $\hat{y_i}$ is 9-bit without removing the mean. When coding a Kodak image whose resolution is 768 $\times$ 512, $\boldsymbol{\hat{y}_k}$ is 48 $\times$ 32 $\times$ 128. Therefore, the bit-depth for the specific channel is 13824-bit ($48 \times 32 \times 9$). After reducing the mean, the largest bit-depth for each element $\hat{y_i}$ can be reduced to 8-bit, thus 12288-bit ($48 \times 32 \times 8$) is consumed for all the $\hat{y_i}$ in the specific channel. In addition, the estimated mean is sent to the decoder which requires 8-bit. Overall, 12296-bit is consumed, which can save about 11\% compared with the origin. For the two higher $\lambda$ models when optimizing MS-SSIM, the largest bit-depth for each element $\hat{y_i}$ can be reduced from 10-bit to 9-bit. Overall, the bit-depth for the specific channel can be reduced from 15360-bit to 13832-bit as shown in Table \ref{table_reduce_mean}. About 10\% bit consumption can be saved.

\begin{table}[t]
	\setlength{\abovecaptionskip}{-0.5cm}   
	\setlength{\belowcaptionskip}{0.0cm}
	\caption{Required bit-width for $\hat{y}$ for the specific channel with a large non-zero mean}
	\footnotesize
	\begin{center}
		{
			\resizebox{0.9\width}{!}{
			
			\begin{tabular}{|c|c|c|c|c|}
				\hline
				\textbf{$\lambda$}					
				&\multicolumn{2}{c}{$0.3, 0.5$}
				&\multicolumn{2}{|c|}{$80, 128$} \\ \hline
				
				\textbf{Component}					
				&\text{$\hat{y}$}	
				&\text{mean}
				&\text{$\hat{y}$}	
				&\text{mean} \\ \hline								
				
				\textbf{Before reducing mean}			
				&\text{9-bit / element}		
				&\text{-}
				&\text{10-bit / element}		
				&\text{-}		\\ \hline				
				
				\textbf{After reducing mean}			
				&\text{8-bit / element}		
				&\text{$8$-bit}
				&\text{9-bit / element}		
				&\text{$8$-bit}		\\ \hline	
				
			\end{tabular}
		
			}
		
		}
	\end{center}
	\label{table_reduce_mean}
	\vspace{-4mm}
\end{table}

\section{Experimental Results}

\subsection{Training and Network Details}

For the training set, we use 256 $\times$ 256 patches cropped from ImageNet \cite{deng2009imagenet}, and set batch size as eight. For the distortion term in the loss function in Eq. \ref{eq_rdcost}, $D(x,\hat{x})$ is MSE when optimizing PSNR, and it is 1-MS-SSIM($x,\hat{x}$) when optimizing the MS-SSIM. Since different $\lambda$ can achieve various trade-off between distortion and rate, we train several models to obtain a R-D curve. When optimizing MSE, $\lambda$ is selected from \{0.001625, 0.00325, 0.0075, 0.015, 0.03, 0.05\}. When optimizing MS-SSIM, $\lambda$ is picked up from \{3, 5, 10, 40, 80, 128\}. We train about $1 \times 10^6$ iterations for all the $\lambda$, the learning rate is set as $1 \times 10^{-4}$ at first and decayed to $1 \times 10^{-5}$ for the final 80K iterations. For the test set, we use 24 lossless Kodak images \cite{franzen1999kodak} that are commonly used in the previous works. The coding gain and bpp (bit per pixel) are evaluated in the next sub-section. Noted that there is no overlap between the training set and the test set.

For the network architecture, we use 128 channels when $\lambda$ is smaller than 0.015, and 192 channels for the other $\lambda$ in the case of MSE optimization. When optimizing MS-SSIM, we use 128 channels for the $\lambda$ smaller than 80, and 192 channels for the remaining ones. For the Leaky-ReLU adopted in the hyper path, there is a fixed coefficient when dealing with the negative values. We set the coefficient as 0.125 to make it friendly for a potential hardware design in the future.

\subsection{Coding Gain Evaluation}

There are three parts for the coding gain evaluation. First, we evaluate the overall coding gain of our proposal compared with the previous works. Second, we evaluate the effect of the proposed WCFT in the weight quantization. Third, we evaluate the effect of the proposed NLQ codebook in the activation quantization.

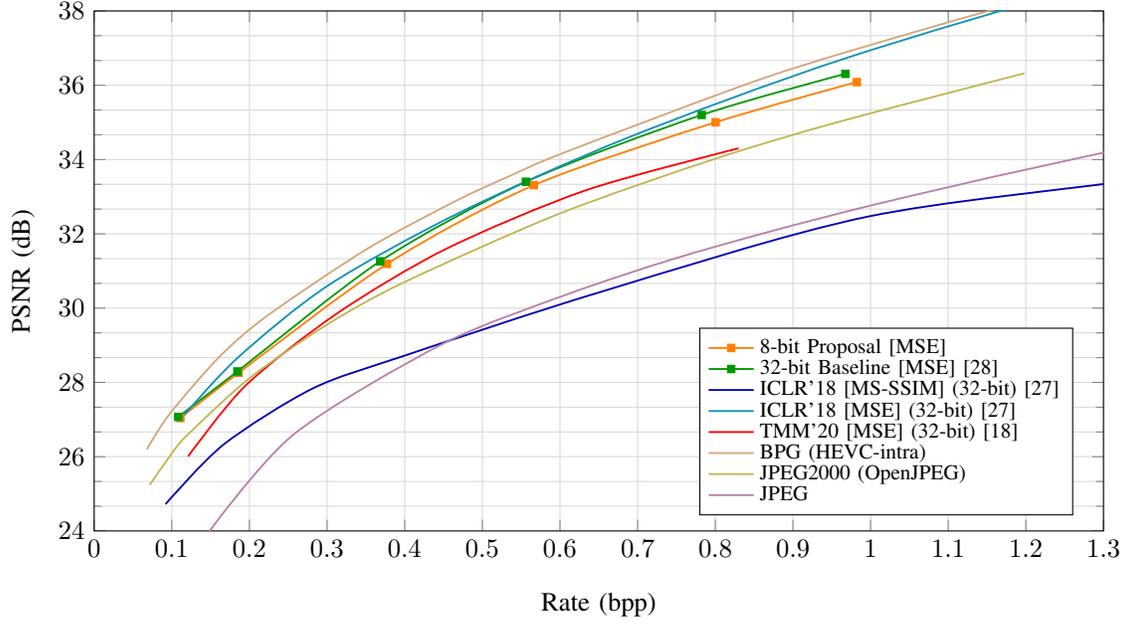
\begin{figure*}
	\centering
	\pgfplotsset{
	}
	\pgfplotsset{every axis plot/.append style={line width=0.7pt}}
	\tikzset{every mark/.append style={scale=0.5}}
	\begin{tikzpicture}
	\begin{axis}[
	width=15cm, height=8.5cm,
	x label style={at={(axis description cs:0.5,-0.1)},anchor=north},
	y label style={at={(axis description cs:-0.05,.5)},anchor=south},
	y tick label style = { /pgf/number format/.cd, precision=3, /tikz/.cd},
	xlabel = {Rate (bpp)},
	ylabel = {PSNR (dB)},
	xmin = 0.00,
	xmax = 1.30,
	ymin = 24.0,
	ymax = 38.0,
	minor y tick num = 2,
	grid = both,
	grid style = {gray!30},
	legend entries = {8-bit Proposal [MSE], 32-bit Baseline [MSE]\cite{cheng2019deep}, ICLR'18 [MS-SSIM] (32-bit) \cite{balle2018variational}, ICLR'18 [MSE] (32-bit)\cite{balle2018variational}, TMM'20 [{MSE}] (32-bit) \cite{Cheng2019EnergyCI} , BPG (HEVC-intra), JPEG2000 (OpenJPEG), JPEG},
	legend style={font=\fontsize{8}{8}\selectfont, row sep=-3pt},
	legend cell align=left,
	legend pos = {south east},
	]
	\addplot[smooth, orange, mark=square*, mark size=2.5pt] table {data_proposal-MSE_PSNR.dat};
	\addplot[smooth, green!60!black, mark=square*, mark size=2.5pt] table {data_baseline-MSE_PSNR.dat};
	\addplot[smooth, blue!70!black] table {data_ICLR18-MSSSIM_PSNR.dat};
	\addplot[smooth, cyan!70!black] table {data_ICLR18-MSE_PSNR.dat};
	\addplot[smooth, red] table {data_TMM20_PSNR.dat};
	\addplot[smooth, brown!70!white] table {data_BPG_PSNR.dat};
	\addplot[smooth, yellow!70!black] table {data_JPEG2000_PSNR.dat};
	\addplot[smooth, DarkOrchid!70!white] table {data_JPEG_PSNR.dat};
	
	\end{axis}
	\end{tikzpicture}
	\caption{PSNR comparison of proposed 8-bit quantized LIC with 32-bit baseline LIC \cite{cheng2019deep}, ICLR'18\cite{balle2018variational}, TMM'20\cite{Cheng2019EnergyCI}, BPG, JPEG2000 and JPEG.}
	\label{fig_psnr}
	
\end{figure*}

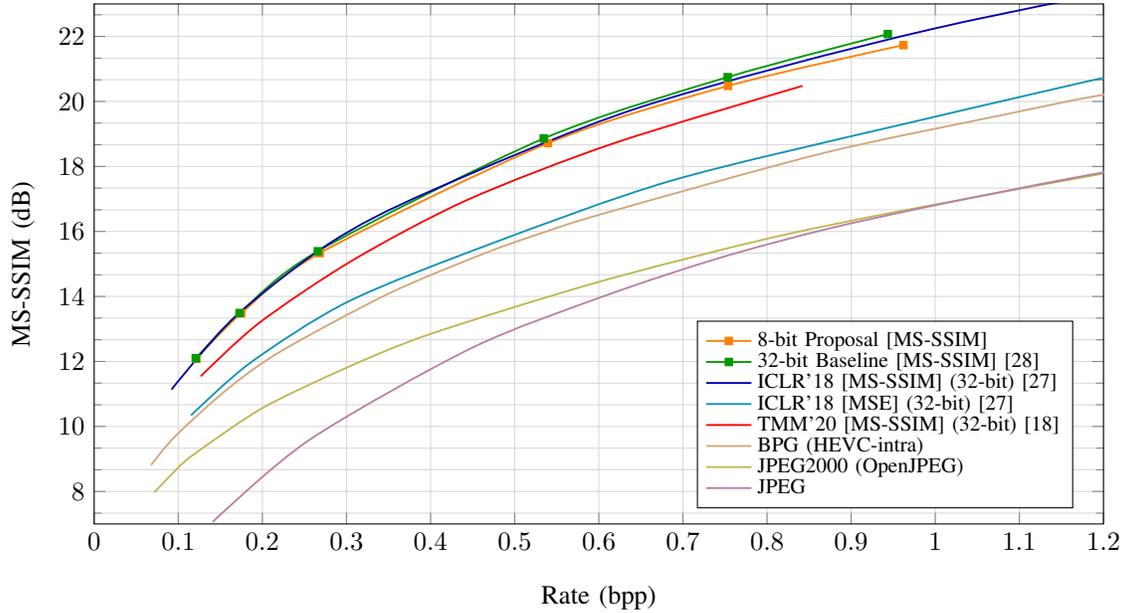
\begin{figure*}
	\centering
	\pgfplotsset{
	}
	\pgfplotsset{every axis plot/.append style={line width=0.7pt}}
	\tikzset{every mark/.append style={scale=0.5}}
	\begin{tikzpicture}
	\begin{axis}[
	width=15cm, height=8.5cm,
	x label style={at={(axis description cs:0.5,-0.1)},anchor=north},
	y label style={at={(axis description cs:-0.05,.5)},anchor=south},
	y tick label style = { /pgf/number format/.cd, precision=3, /tikz/.cd},
	xlabel = {Rate (bpp)},
	ylabel = {MS-SSIM (dB)},
	xmin = 0.00,
	xmax = 1.20,
	ymin = 7.0,
	ymax = 23.0,
	minor y tick num = 2,
	grid = both,
	grid style = {gray!30},
	legend entries = {8-bit Proposal [MS-SSIM], 32-bit Baseline [MS-SSIM]\cite{cheng2019deep}, ICLR'18 [MS-SSIM] (32-bit) \cite{balle2018variational}, ICLR'18 [MSE] (32-bit)\cite{balle2018variational}, TMM'20 [MS-SSIM] (32-bit) \cite{Cheng2019EnergyCI} , BPG (HEVC-intra), JPEG2000 (OpenJPEG), JPEG},
	legend style={font=\fontsize{8}{8}\selectfont, row sep=-3pt},
	legend cell align=left,
	legend pos = {south east},
	]
	\addplot[smooth, orange, mark=square*, mark size=2.5pt] table {data_proposal-MSSSIM_MSSSIM.dat};
	\addplot[smooth, green!60!black, mark=square*, mark size=2.5pt] table {data_baseline-MSSSIM_MSSSIM.dat};
	\addplot[smooth, blue!70!black] table {data_ICLR18-MSSSIM_MSSSIM.dat};
	\addplot[smooth, cyan!70!black] table {data_ICLR18-MSE_MSSSIM.dat};
	\addplot[smooth, red] table {data_TMM20_MSSSIM.dat};
	\addplot[smooth, brown!70!white] table {data_BPG_MSSSIM.dat};
	\addplot[smooth, yellow!70!black] table {data_JPEG2000_MSSSIM.dat};
	\addplot[smooth, DarkOrchid!70!white] table {data_JPEG_MSSSIM.dat};
	
	\end{axis}
	\end{tikzpicture}
	\caption{MS-SSIM comparison of proposed 8-bit quantized LIC with 32-bit baseline LIC \cite{cheng2019deep}, ICLR'18\cite{balle2018variational}, TMM'20\cite{Cheng2019EnergyCI}, BPG, JPEG2000 and JPEG.}
	\label{fig_ssim}
	
\end{figure*}

First, the overall results are shown in Fig. \ref{fig_psnr} and Fig. \ref{fig_ssim}. For the 32-bit floating point baseline \cite{cheng2019deep}, we can achieve a comparable performance with \cite{balle2018variational}. In fact, compared with \cite{balle2018variational}, we have two major differences. One is on the hyper path, we not only use the hyper path to estimate the scale, but also estimate the mean, which can improve the performance. The other difference is on the activation layer. We simply use ReLU in the main path to reduce the complexity. As a result, the coding gain reaches the same level with \cite{balle2018variational}. Based on the baseline, we quantize each weight element and activation element from 32-bit floating-point to 8-bit fixed-point. The coding performance degradation coming from the quantization is small as shown in Table \ref{table_overall}. When only quantizing the weights or activations, the coding gain loss is less than 0.2dB in the worst case. If we quantize both weights and activations, there is at most 0.339dB coding loss for the largest $\lambda$ in the MS-SSIM case. From Table \ref{table_overall}, we can also observe that the coding loss for the larger $\lambda$ is large. It is because the values of both weights and activations are expanded for large $\lambda$. As a result, at the constraint of the same bit-budget, the QP for large $\lambda$ is degraded. We also compare with \cite{Cheng2019EnergyCI} and three traditional standards that are JPEG, JPEG2000 and BPG. For the MS-SSIM, the quantized version can outperform BPG and \cite{Cheng2019EnergyCI} significantly. For the PSNR, the quantized version is better than JPEG2000 and \cite{Cheng2019EnergyCI}.

Second, the effects of using WCFT is shown in Table \ref{table_WCFT}. As shown in the line 5 of the pseudo code \textbf{Algorithm 1}, we clip weights in several rounds $I_2$. In our experiments, for both MSE optimized and MS-SSIM optimized cases, the clipping is performed for one round and three rounds for three lower $\lambda$ and three higher $\lambda$, respectively. For the three lower $\lambda$, $\beta$ is set as one and $I_3$ is about $1.4 \times 10^5$ fine tuning iterations. For the three higher $\lambda$, $\beta$ is set as 1, $\sqrt{2}$ and 1 for three rounds, respectively. $I_3$ is about $8.6 \times 10^4$, $8.6 \times 10^4$ and $1.4 \times 10^5$ for three rounds, respectively. Before using WCFT, the coding loss increases with larger $\lambda$, and it is as large as 0.756dB in the case of $\lambda$ being 128. After using WCFT, the coding loss can be reduced for all the $\lambda$ significantly especially for large $\lambda$. In the case of $\lambda$ being 128, the coding loss can be reduced from 0.756dB to 0.185dB.

Third, the effects of the proposed NLQ codebook is shown in Table \ref{table_output}. When using LQ for the activation, the quantization range is not fully utilized thus the coding loss is not so small. The trend of the coding loss is similar with the weight quantization, which will increase with large $\lambda$. After using the NLQ, the coding loss can be reduced for almost all the cases. Especially, for the $\lambda$ being 128, the coding loss can be reduced from 0.297dB to 0.159dB.

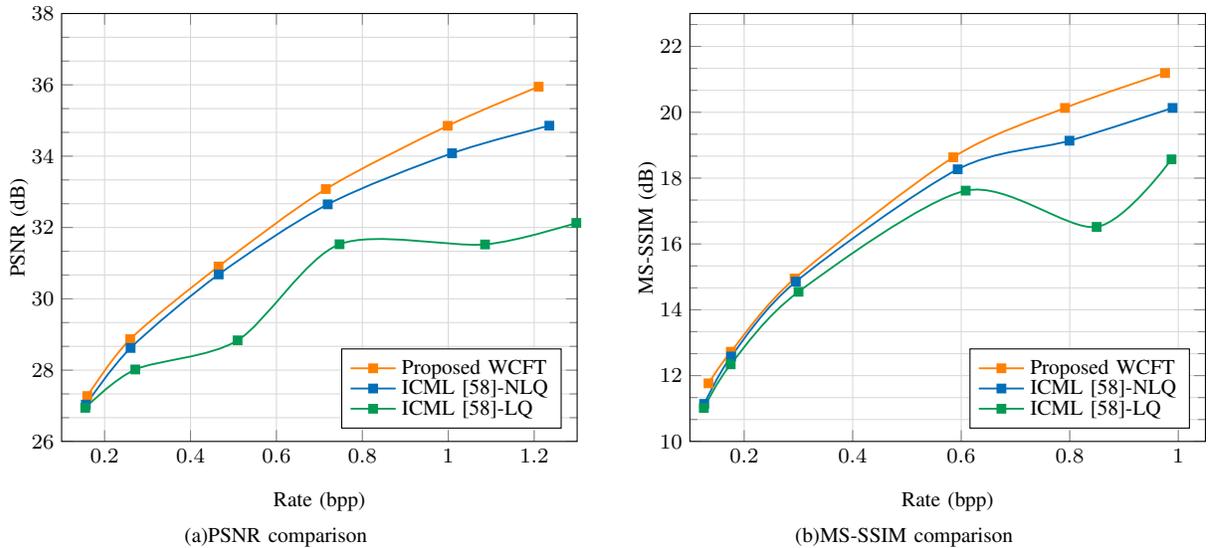
\begin{figure*}[h]
	\footnotesize
	\centering
	
	\begin{minipage}[b]{0.4\linewidth}
		\pgfplotsset{
		}
		\pgfplotsset{every axis plot/.append style={line width=0.7pt}}
		\tikzset{every mark/.append style={scale=0.5}}
		\begin{tikzpicture}
		\begin{axis}[
		x label style={at={(axis description cs:0.5,-0.1)},anchor=north},
		y label style={at={(axis description cs:-0.05,.5)},anchor=south},
		y tick label style = { /pgf/number format/.cd, precision=3, /tikz/.cd},
		xlabel = {Rate (bpp)},
		ylabel = {PSNR (dB)},
		xmin = 0.10,
		xmax = 1.30,
		ymin = 26.0,
		ymax = 38.0,
		minor y tick num = 2,
		grid = both,
		grid style = {gray!30},
		legend entries = {Proposed WCFT, ICML\cite{lin2016fixed}-NLQ,
			ICML\cite{lin2016fixed}-LQ},
		legend style={font=\fontsize{8}{8}\selectfont, row sep=-3pt},
		legend cell align=left,
		legend pos = {south east},
		]
		\addplot[smooth, orange, mark=square*, mark size=3.0pt] table {r2q2-wcft-quanw-psnr.dat};
		\addplot[smooth, NavyBlue, mark=square*, mark size=3.0pt] table {r2q2-gaussian-nlq-psnr.dat};
		\addplot[smooth, ForestGreen, mark=square*, mark size=3.0pt] table {r2q2-gaussian-x3-lq-psnr.dat};
		
		\end{axis}
		\end{tikzpicture}
		
		\centerline{(a)PSNR comparison}
		
	\end{minipage}\hspace{10mm}
	\begin{minipage}[b]{0.4\linewidth}
		\pgfplotsset{
		}
		\pgfplotsset{every axis plot/.append style={line width=0.7pt}}
		\tikzset{every mark/.append style={scale=0.5}}
		\begin{tikzpicture}
		\begin{axis}[
		x label style={at={(axis description cs:0.5,-0.1)},anchor=north},
		y label style={at={(axis description cs:-0.05,.5)},anchor=south},
		y tick label style = { /pgf/number format/.cd, precision=3, /tikz/.cd},
		xlabel = {Rate (bpp)},
		ylabel = {MS-SSIM (dB)},
		xmin = 0.10,
		xmax = 1.05,
		ymin = 10.0,
		ymax = 23.0,
		minor y tick num = 2,
		grid = both,
		grid style = {gray!30},
		legend entries = {Proposed WCFT, ICML\cite{lin2016fixed}-NLQ,
			ICML\cite{lin2016fixed}-LQ},
		legend style={font=\fontsize{8}{8}\selectfont, row sep=-3pt},
		legend cell align=left,
		legend pos = {south east},
		]
		\addplot[smooth, orange, mark=square*, mark size=3.0pt] table {r2q2-wcft-quanw-msssim.dat};
		\addplot[smooth, NavyBlue, mark=square*, mark size=3.0pt] table {r2q2-gaussian-nlq-msssim.dat};
		\addplot[smooth, ForestGreen, mark=square*, mark size=3.0pt] table {r2q2-gaussian-x3-lq-msssim.dat};
		
		\end{axis}
		\end{tikzpicture}
		
		\centerline{(b)MS-SSIM comparison}
		
	\end{minipage}

	\caption{Coding gain comparison of quantization methods.}
	\label{fig_qualcomm}
	
\end{figure*}

{
In addition to the hyperprior model, we also apply our methods to factorized-prior model. The main path of the factorized-prior model is the same as hyper model, which is to use four layers in both analysis and synthesis transform. We train 6 models for MSE and MS-SSIM, respectively. The setting of $\lambda$ is exactly same as hyper model. For the lower three $\lambda$, 128 channels are used. For the higher three $\lambda$, 192 channels are used.

We evaluate the coding loss of our method compared with the floating-point factorized-prior model. First, we give the overall coding performance in Table \ref{table_overall}. When only quantizing weights or activations, the coding loss is within 0.2dB. When quantizing both weight and activation, the largest coding loss is 0.229dB. We can observe that the coding loss is large for large $\lambda$, which is similar as hyperprior model. Second, we evaluate the effect of proposed WCFT in Table \ref{table_WCFT}. The setting of $\beta$ and $I_3$ in the fine tuning is the same as hyperprior model. As a result, before using WCFT, the coding loss is as large as 0.716dB in the case of $\lambda$ being 0.05. After using WCFT, the coding loss can be significantly reduced to 0.166dB. Finally, we evaluate the effect of proposed NLQ for activation quantization in Table \ref{table_output}. We can see that the coding loss can be reduced for almost all the $\lambda$ by using the proposed NLQ.

To evaluate the efficiency of our quantization method, we compare with \cite{lin2016fixed} which also focused on the fixed-point quantization. In \cite{lin2016fixed}, there are three steps to decide the number of fractional bits. The first step is to determine the effective standard deviation of the quantity being quantized $\xi$ in the following equation
\begin{equation}
\xi = \alpha \cdot \sigma
\label{eq_qualcomm_1}
\end{equation}
where $\sigma$ is the standard deviation of the Gaussian distribution and $\alpha$ is set as 3 due to the longer tails than Gaussian. The second step is to calculate step size by the following equation
\begin{equation}
s=\xi \cdot Stepsize(\beta)
\label{eq_qualcomm_2}
\end{equation}
where $\beta$ is the bitdepth and $Stepsize(\beta)$ is the step-size of optimal symmetric uniform quantizer. The third step is to compute number of fractional bits as shown in the following equation.
\begin{equation}
n=-\left \lceil{log_2s} \right \rceil
\label{eq_qualcomm_3}
\end{equation}
We applied the above method in the weight quantization and compare with our proposed WCFT. For Eq. \ref{eq_qualcomm_1}, $\sigma$ is estimated for each channel. For the Stepsize in Eq. \ref{eq_qualcomm_2}, it is equal to 0.0308 in the case of 8-bit budget. To determine $n$ in Eq. \ref{eq_qualcomm_3}, we follow \cite{lin2016fixed} that is to ensure $2^{-n}$ as the highest precision of the fixed point representation. The results are shown in Fig. \ref{fig_qualcomm}. We can see that the R-D curve of \cite{lin2016fixed} is not smooth for different $\lambda$. The reason is that the uniform quantization cannot always ensure adequate large dynamic range and  high quantization precision at the same time. Therefore, we also apply the non-uniform quantization based on the number of fractional bits in \cite{lin2016fixed}. We use the same NLQ codebook for our method and \cite{lin2016fixed} which is shown in the below
\begin{equation}
\small
Q(w) = \begin{cases}
\dfrac{ round( w\times2^{n-3} )  }{2^{n-3}}   , & \text{if } |w|\in[ \frac{64}{2^{n}}+\frac{16}{2^{n-2}}\\ &\frac{64}{2^{n}}+\frac{16}{2^{n-2}}+\frac{48}{2^{n-3}})\\
\dfrac{ round( w\times2^{n-2} )  }{2^{n-2}}  , & \text{if } |w|\in[\frac{64}{2^{n}}, \frac{64}{2^{n}}+\frac{16}{2^{n-2}})\\
\dfrac{ round( w\times2^{n} )  }{2^{n}} , & \text{if } |w|\in[0, \frac{64}{2^{n}})\\
\end{cases}
\label{eq_qualcomm_4}
\end{equation}
where $w$ and $Q(w)$ are the weights before and after the quantization. Considering the non-uniform quantization could reach a large dynamic range, we set $\alpha$ as 1.5 to ensure a higher precision. The results are shown in Fig. \ref{fig_qualcomm}. From the results, we can see that we can achieve better coding gain than \cite{lin2016fixed} for all the $\lambda$, which illustrates the effect of proposed WCFT.

}

\begin{table*}[t]
	\setlength{\abovecaptionskip}{0.0cm}   
	\setlength{\belowcaptionskip}{0.0cm}
	\caption{Coding loss compared with the baseline (32-bit floating version) when quantizing weight and activation}
	\scriptsize
	\begin{center}
		{
			
			\resizebox{0.85\width}{!}{
			
			\begin{tabular}{|c|c|c|c|c|c|c|c|c|c|c|c|c|c|c| } \hline
				
				\text{}  
				&\text{}
				&\text{}
				&\multicolumn{6}{c}{\text{MSE optimized}}  	
				&\multicolumn{6}{|c|}{\text{MS-SSIM optimized}}  \\ \hline	
				
				\textbf{}  
				&\textbf{}
				&$\lambda$ 
				&\text{$0.001625$}		
				&\text{$0.00325$}			
				&\text{$0.0075$}		
				&\text{$0.015$}		
				&\text{$0.03$}		
				&\text{$0.05$}			
				&\text{$3$}		
				&\text{$5$}			
				&\text{$10$}		
				&\text{$40$}		
				&\text{$80$}		
				&\text{$128$}			\\ \hline
				
				\multirow{6}*{\textbf{Hyper}} 
				&\textbf{Quan.}
				&\text{PSNR/MS-SSIM (dB)}					
				&{$-0.009$}		
				&{$-0.031$}		
				&{$-0.043$}		
				&{$-0.068$}		
				&{$-0.142$}		
				&{$-0.147$}		
				&{$-0.007$}		
				&{$+0.009$}		
				&{$-0.034$}		
				&{$-0.087$}	
				&{$-0.167$}		
				&{$-0.185$} 	\\ 					
					
				\textbf{}
				&\textbf{weight} 
				&\text{bpp}					
				&$0.001$		
				&$0.001$		
				&$0.006$		
				&$0.002$		
				&$0.004$		
				&$0.006$
				&$0.001$		
				&$0.002$	
				&$0.002$	
				&$0.003$		
				&$-0.003$		
				&$0.014$	\\ \cline{2-15}

				\textbf{}
				&\textbf{Quan.}
				&\text{PSNR/MS-SSIM (dB)}
				&{$-0.020$}		
				&{$-0.006$}		
				&{$-0.028$}		
				&{$-0.025$}		
				&{$-0.053$}		
				&{$-0.077$}		
				&{$-0.008$}		
				&{$-0.015$}		
				&{$-0.026$}		
				&{$-0.058$}	
				&{$-0.109$}		
				&{$-0.159$} 	\\ 
				
				\textbf{}
				&\textbf{acti} 
				&\text{bpp}
				&$0.002$		
				&$0.001$		
				&$0.003$		
				&$0.010$		
				&$0.012$		
				&$0.010$
				&$0.0003$		
				&$0.0004$	
				&$0.001$	
				&$0.002$		
				&$0.003$		
				&$0.004$	\\ \cline{2-15}				
				
				\textbf{}
				&\textbf{Quan.}
				&\text{PSNR/MS-SSIM (dB)}
				&\text{$-0.031$}		
				&\text{$-0.037$}		
				&\text{$-0.070$}		
				&\text{$-0.091$}		
				&\text{$-0.198$}		
				&\text{$-0.222$}		
				&\textbf{$-0.017$}		
				&\textbf{$-0.005$}		
				&\textbf{$-0.057$}		
				&\textbf{$-0.145$}	
				&\text{$-0.271$}		
				&\text{$-0.339$} 	\\ 
				
				\textbf{}
				&\textbf{weight+acti} 
				&\text{bpp}
				&\text{$0.003$}		
				&\text{$0.002$}		
				&\text{$0.009$}		
				&\text{$0.010$}		
				&\text{$0.018$}		
				&\text{$0.015$}
				&\textbf{$0.001$}		
				&\textbf{$0.002$}		
				&\textbf{$0.002$}		
				&\textbf{$0.005$}		
				&\text{$0.0003$}		
				&\text{$0.019$}	\\ \hline
				
				\textbf{}
				\textbf{}
				&\textbf{Quan.}
				&\text{PSNR/MS-SSIM (dB)}
				&$-0.014$
				&$-0.017$		
				&$-0.084$		
				&$-0.071$		
				&$-0.152$		
				&$-0.166$
				&$-0.020$		
				&$-0.033$	
				&$-0.041$		
				&$-0.032$	
				&$-0.090$		
				&$-0.116$ 	\\ 
				
				\textbf{}
				&\textbf{weight} 
				&\text{bpp}
				&$0.0001$	
				&$0.004$		
				&$0.005$
				&$0.003$		
				&$0.002$		
				&$0.002$
				&$-0.0002$	
				&$-0.0004$		
				&$-0.001$		
				&$0.002$		
				&$0.001$		
				&$0.003$	\\ \cline{2-15}
				
				\textbf{Facto-}
				&\textbf{Quan.}
				&\text{PSNR/MS-SSIM (dB)}
				&$-0.007$		
				&$-0.009$		
				&$-0.015$		
				&$-0.020$		
				&$-0.040$		
				&$-0.061$
				&$-0.013$		
				&$-0.016$	
				&$-0.014$		
				&$-0.032$	
				&$-0.090$		
				&$-0.116$ 	\\ 
				
				\textbf{rized}
				&\textbf{acti} 
				&\text{bpp}
				&$0.0001$	
				&$0.0001$		
				&$0.0004$
				&$0.001$		
				&$0.001$		
				&$0.001$
				&$0.000$	
				&$0.0001$
				&$0.000$		
				&$0.0004$		
				&$0.001$		
				&$0.001$	\\ \cline{2-15}

				\textbf{}
				&\textbf{Quan.}
				&\text{PSNR/MS-SSIM (dB)}
				&$-0.020$		
				&$-0.025$		
				&$-0.098$		
				&$-0.091$		
				&$-0.194$		
				&$-0.224$
				&$-0.033$		
				&$-0.049$	
				&$-0.068$		
				&$-0.063$	
				&$-0.179$		
				&$-0.229$ 	\\

				\textbf{}
				&\textbf{weight+acti}
				&\text{bpp}
				&$0.0002$	
				&$0.004$		
				&$0.005$
				&$0.004$		
				&$0.003$		
				&$0.003$
				&$-0.0002$	
				&$-0.0003$
				&$-0.001$		
				&$0.002$		
				&$0.001$		
				&$0.004$	\\ \hline			
				
			\end{tabular}
		
			}
		}
	\end{center}
	\label{table_overall}
	\vspace{-6mm}
\end{table*}

\begin{table*}[t]
	\setlength{\abovecaptionskip}{0.0cm}   
	\setlength{\belowcaptionskip}{0.0cm}
	\caption{Coding gain improvement by using proposed WCFT for weight quantization}
	\scriptsize
	\begin{center}
		{
			
			\resizebox{0.85\width}{!}{
				
				\begin{tabular}{|c|c|c|c|c|c|c|c|c|c|c|c|c|c|c|c|} \hline
					
					\textbf{}  
					&\text{}
					&\text{}
					&\multicolumn{6}{c}{\text{MSE optimized}}  	
					&\multicolumn{6}{|c|}{\text{MS-SSIM optimized}}  \\ \hline		
					
					\textbf{} 
					&\text{}
					&$\lambda$ 
					&{$0.001625$}		
					&{$0.00325$}			
					&{$0.0075$}		
					&{$0.015$}		
					&{$0.03$}		
					&{$0.05$}			
					&{$3$}		
					&{$5$}			
					&{$10$}		
					&{$40$}		
					&{$80$}		
					&{$128$}			\\ \hline
					
					\multirow{4}*{\textbf{Hyper}} 
					&\textbf{W/o} 	
					&\text{PSNR/MS-SSIM (dB)}
					&\text{$-0.030$}		
					&\text{$-0.039$}		
					&\text{$-0.231$}		
					&\text{$-0.280$}		
					&\text{$-0.429$}		
					&\text{$-0.625$}		
					&\text{$-0.021$}		
					&\text{$-0.037$}		
					&\text{$-0.057$}		
					&\text{$-0.262$}		
					&\text{$-0.540$}		
					&\text{$-0.756$}		\\ 
					
					&\text{}
					\textbf{WCFT}
					&\text{bpp}
					&\text{$0.001$}		
					&\text{$0.005$}		
					&\text{$0.008$}		
					&\text{$0.009$}		
					&\text{$0.028$}		
					&\text{$0.021$} 
					&\text{$0.001$}		
					&\text{$0.001$}		
					&\text{$0.001$}		
					&\text{$0.015$}		
					&\text{$0.011$}		
					&\text{$0.026$}			\\ \cline{2-15}
					
				    \text{}
				    &\textbf{W/} 
					&\text{PSNR/MS-SSIM (dB)}				
					&{$-0.009$}		
					&{$-0.031$}		
					&{$-0.043$}		
					&{$-0.068$}		
					&{$-0.142$}		
					&{$-0.147$}		
					&{$-0.007$}		
					&{$+0.009$}		
					&{$-0.034$}		
					&{$-0.087$}	
					&{$-0.167$}		
					&{$-0.185$} 	\\ 							
					
					\text{}
					&\textbf{WCFT}
					&\text{bpp}			
					&$0.001$		
					&$0.001$		
					&$0.006$		
					&$0.002$		
					&$0.004$		
					&$0.006$
					&$0.001$		
					&$0.002$	
					&$0.002$	
					&$0.003$		
					&$-0.003$		
					&$0.014$	\\ \hline
					
					\text{}
					&\textbf{W/o}
					&\text{PSNR/MS-SSIM (dB)}
					&$-0.012$		
					&$-0.119$		
					&$-0.279$		
					&$-0.145$		
					&$-0.485$		
					&$-0.716$		
					&$-0.013$		
					&$-0.041$		
					&$0.000$		
					&$-0.250$		
					&$-0.597$		
					&$-0.712$		\\ 
					
					\textbf{Facto-}
					&\textbf{WCFT} 
					&\text{bpp}
					&$0.000$		
					&$0.001$		
					&$0.001$		
					&$0.006$		
					&$0.012$		
					&$0.010$ 
					&$0.001$		
					&$0.000$		
					&$0.003$		
					&$0.005$		
					&$0.001$		
					&$0.009$			\\ \cline{2-15}
					
					\textbf{rized}
					&\textbf{W/}
					&\text{PSNR/MS-SSIM (dB)}					
					&$-0.014$		
					&$-0.017$		
					&$-0.084$		
					&$-0.071$		
					&$-0.152$		
					&$-0.166$
					&$-0.020$		
					&$-0.033$	
					&$-0.041$		
					&$-0.032$	
					&$-0.090$		
					&$-0.116$ 	\\ 					
					
					\text{}
					&\textbf{WCFT} 
					&\text{bpp}					
					&$0.0001$	
					&$0.004$		
					&$0.005$
					&$0.003$		
					&$0.002$		
					&$0.002$
					&$-0.0002$	
					&$-0.0004$		
					&$-0.001$		
					&$0.002$		
					&$0.001$		
					&$0.003$	\\ \hline

				\end{tabular}
				
			}
		}
	\end{center}
	\label{table_WCFT}
	\vspace{-6mm}
\end{table*}

\begin{table*}[t]
	\setlength{\abovecaptionskip}{0.0cm}   
	\setlength{\belowcaptionskip}{0.0cm}
	\caption{Coding gain improvement by using proposed NLQ codebook for activation quantization}
	\scriptsize
	\begin{center}
		{
			
			\resizebox{0.85\width}{!}{
				
				\begin{tabular}{|c|c|c|c|c|c|c|c|c|c|c|c|c|c|c|c|} \hline
					
					\textbf{}
					&\textbf{}
					&\text{}  
					&\multicolumn{6}{c}{\text{MSE optimized}}  	
					&\multicolumn{6}{|c|}{\text{MS-SSIM optimized}}  \\ \hline		
					
					\textbf{}
					&\textbf{}
					&$\lambda$ 
					&{$0.001625$}		
					&{$0.00325$}			
					&{$0.0075$}		
					&{$0.015$}		
					&{$0.03$}		
					&{$0.05$}			
					&{$3$}		
					&{$5$}			
					&{$10$}		
					&{$40$}		
					&{$80$}		
					&{$128$}			\\ \hline
					
					\multirow{4}*{\textbf{Hyper}}
					&\textbf{W/o}
					&\text{PSNR/MS-SSIM (dB)}
					&\text{$-0.019$}		
					&\text{$-0.009$}		
					&\text{$-0.042$}		
					&\text{$-0.051$}		
					&\text{$-0.131$}		
					&\text{$-0.161$}		
					&\text{$-0.021$}		
					&\text{$-0.027$}		
					&\text{$-0.046$}		
					&\text{$-0.106$}		
					&\text{$-0.196$}		
					&\text{$-0.297$}		\\
					
					\textbf{}
					&\textbf{NLQ} 
					&\text{bpp}
					&\text{$0.002$}		
					&\text{$0.001$}		
					&\text{$0.003$}		
					&\text{$0.011$}		
					&\text{$0.015$}		
					&\text{$0.013$} 
					&\text{$0.0003$}		
					&\text{$0.0004$}		
					&\text{$0.001$}		
					&\text{$0.002$}		
					&\text{$0.003$}		
					&\text{$0.005$}			\\ \cline{2-15}
					
					\textbf{}
					&\textbf{W/}
					&\text{PSNR/MS-SSIM (dB)}
					&{$-0.020$}		
					&{$-0.006$}		
					&{$-0.028$}		
					&{$-0.025$}		
					&{$-0.053$}		
					&{$-0.077$}		
					&{$-0.008$}		
					&{$-0.015$}		
					&{$-0.026$}		
					&{$-0.058$}	
					&{$-0.109$}		
					&{$-0.159$} 	\\ 				

					\textbf{}
					&\textbf{NLQ} 
					&\text{bpp}
					&$0.002$		
					&$0.001$		
					&$0.003$		
					&$0.010$		
					&$0.012$		
					&$0.010$
					&$0.0003$		
					&$0.0004$	
					&$0.001$	
					&$0.002$		
					&$0.003$		
					&$0.004$	\\ \hline	
					
					\textbf{}
					&\textbf{W/o}
					&\text{PSNR/MS-SSIM (dB)}
					&$-0.008$		
					&$-0.014$		
					&$-0.031$		
					&$-0.037$		
					&$-0.082$		
					&$-0.142$						
					&$-0.020$		
					&$-0.025$		
					&$-0.054$		
					&$-0.095$		
					&$-0.179$		
					&$-0.284$	\\

					\textbf{Facto-}
					&\textbf{NLQ} 
					&\text{bpp}
					&$0.0001$		
					&$0.0002$		
					&$0.001$		
					&$0.001$		
					&$0.002$		
					&$0.002$ 
					&$0.000$	
					&$0.0001$		
					&$0.0001$		
					&$0.001$		
					&$0.001$		
					&$0.001$			\\ \cline{2-15}
					
					\textbf{rized}
					&\textbf{W/}
					&\text{PSNR/MS-SSIM (dB)}
					&$-0.007$		
					&$-0.009$		
					&$-0.015$		
					&$-0.020$		
					&$-0.040$		
					&$-0.061$
					&$-0.013$		
					&$-0.016$	
					&$-0.014$		
					&$-0.032$	
					&$-0.090$		
					&$-0.116$ 	\\ 	
					
					\textbf{}
					&\textbf{NLQ}
					&\text{bpp}					
					&$0.0001$
					&$0.0001$		
					&$0.0004$
					&$0.001$		
					&$0.001$		
					&$0.001$
					&$0.000$	
					&$0.0001$
					&$0.000$		
					&$0.0004$		
					&$0.001$		
					&$0.001$	\\ \hline

				\end{tabular}
				
			}
		}
	\end{center}
	\label{table_output}
	\vspace{-6mm}
\end{table*}

\subsection{Memory Consumption Evaluation}

\begin{table*}[t]
	\setlength{\abovecaptionskip}{0.0cm}   
	\setlength{\belowcaptionskip}{0.0cm}
	\caption{Memory Cost Consumption (MB) Comparison for the weight and activation storage}
	\scriptsize
	\begin{center}
		{
			
			\resizebox{0.85\width}{!}{
			
			\begin{tabular}{|c|c|c|c|c|c|c|c|c|c|c|c|c|c|}
				\hline
				\textbf{Codec}							
				&\multicolumn{6}{c}{\text{LIC Encoder}}		
				&\multicolumn{6}{|c|}{\text{LIC Decoder}}			\\ \hline					
				
				\multirow{2}*{\textbf{Component}} 		
				&\multicolumn{3}{c}{\text{Weight}}		
				&\multicolumn{3}{|c|}{\text{Activation}}
				&\multicolumn{3}{c}{\text{Weight}}		
				&\multicolumn{3}{|c|}{\text{Activation}}			\\ \cline{2-13}						
				
				\textbf{}						
				&\text{$w$}		
				&\text{$sf_w$}	
				&\text{total}	
				&\text{$o$}		
				&\text{$sf_o$}	
				&\text{total}
				&\text{$w$}		
				&\text{$sf_w$}	
				&\text{total}	
				&\text{$o$}		
				&\text{$sf_o$}	
				&\text{total}	\\ \hline				
				
				\textbf{ICLR'18 \cite{balle2018variational}}				
				&\text{$14.09$}		
				&\text{$-$}	
				&\text{$14.09 {\color{blue} (-72.03\%) }  $}		
				&\text{$70.43$}		
				&\text{$-$}	
				&\text{$70.43 {\color{blue} (-71.93\%) } $}	
				&\text{$9.93$}		
				&\textit{$-$}	
				&\text{$9.93 {\color{blue} (-70.04\%) } $}	
				&\text{$72.94$}		
				&\text{$-$}	
				&\text{$72.94 {\color{blue} (-72.23\%) } $}	\\ \hline			
				
				\textbf{TMM'20 \cite{Cheng2019EnergyCI}}						
				&\text{$3.77$}		
				&\text{$-$}	
				&\text{$3.77 { (+4.48\%) } $}	
				&\text{$532.81$}		
				&\text{$-$}	
				&\text{$532.81 {\color{blue} (-96.29\%) } $}
				&\text{$3.77$}		
				&\text{$-$}	
				&\text{$3.77 {\color{blue} (-21.14\%) } $}	
				&\text{$338.17$}		
				&\text{$-$}	
				&\text{$338.17 {\color{blue} (-94.01\%) } $}	\\ \hline
				
				\textbf{Baseline  \cite{cheng2019deep}}						
				&\text{$15.77$}		
				&\text{$-$}	
				&\text{$15.77  {\color{blue} (-74.99\%) } $}	
				&\text{$79.09$}		
				&\text{$-$}	
				&\text{$79.09  {\color{blue} (-75.00\%) } $}
				&\text{$11.90$}		
				&\text{$-$}	
				&\text{$11.90  {\color{blue} (-74.99\%) } $}	
				&\text{$81.99$}		
				&\text{$-$}	
				&\text{$81.99  {\color{blue} (-75.00\%) } $}	\\ \hline								
				
				\textbf{Proposal}						
				&\text{$3.942$}		
				&\text{$0.0014$}	
				&\text{$3.943$}	
				&\text{$19.771$}		
				&\text{$0.0018$}	
				&\text{$19.773$}
				&\text{$2.975$}		
				&\text{$0.0012$}	
				&\text{$2.976$}	
				&\text{$20.251$}		
				&\text{$0.0014$}	
				&\text{$20.252$}	\\ \hline		
				
			\end{tabular}}
		}
	\end{center}
	\label{table_memory}
	\vspace{-6mm}
\end{table*}

The memory cost of weight and activation will be apparently reduced after the quantization. However, there is some overhead such as storing the scaling factor. Therefore, we measure the overall cost in this section. The measurement is conducted for the encoder and decoder, respectively. The encoder is composed of $g_a$, $h_a$ and $h_s$, while the decoder is built by $h_s$ and $g_s$. The number of channel is set as 128, and the input image resolution is $768 \times 512$ in the measurement. The results are shown in Table \ref{table_memory}.

We measure the memory cost of weight at first. Originally, for the un-quantized baseline, each weight element is 32-bit. So the overall memory consumption for the weight can be calculated in Eq. \ref{eq_original_weight}
\begin{equation}
\small
\textbf{Original weight cost}
=\sum_{i=0} I_i \times H_i  \times W_i \times O_i \times 4
\label{eq_original_weight}
\end{equation}
\noindent where $i$ is the index of the layer, $I$, $O$, $H$, $W$ are input channel number, output channel number, kernel height ant kernel width, and 4 represents four bytes for one element. There are overall 13 and 10 layers for the encoder and decoder, respectively.

After quantizing each weight from 32-bit floating-point to 8-bit fixed-point, the storage for the weight will become the quarter. About the scaling factor $\mathit{sf}$, since it is limited to the power of two, after going through all the cases, we find 4-bit is enough. In the case of CW grouping, the number of $\mathit{sf}$ is equal to the number of output channels. Therefore, the overall memory consumption can be obtained by Eq. \ref{eq_proposed_weight}
\begin{equation}
\small
\begin{aligned}
\textbf{Proposed weight cost} &=\sum_{i=0} I_i \times H_i  \times W_i \times O_i \times 1 \\
&+\sum_{i=0}O_i \times \tfrac{4}{8}
\end{aligned}
\label{eq_proposed_weight}
\end{equation}
\noindent where $\frac{4}{8}$ represents the half byte memory cost for each $\mathit{sf}$.

About the activation, since most hardware accelerators run the processing of a CNN layer-by-layer as described in \cite{chen2016eyeriss}, the original memory consumption can be calculated by
\begin{equation}
\small
\textbf{Original activation cost}
=\sum_{i=0} OH_i  \times OW_i \times O_i \times 4
\label{eq_origin_o}
\end{equation}
\noindent where $OH_i$, $OW_i$ and $O_i$ are the height, width and number of output feature maps for each layer. After quantizing the activation, there are two storage overheads. One overhead is the scaling factor ($sf_o$) which is the second term in Eq. \ref{eq_quan_o}. $sf_o$ is channel-wise and we find that 4-bit is enough for all the cases. Noted that for the encoder, there is no $sf_o$ for the final layer of the AT since each result of AT is rounded to the integer directly. For the decoder, there is no $sf_o$ for the final layer of ST since the outputs are rounded in the element-wise as the reconstructed pixels. The other overhead is the selection signal for the multiplexer which is only required for the channels in the main path. There are four selections thus 2-bit is required for each multiplexer. The overhead can be calculated as the third term in the following equation.
\begin{equation}
\small
\begin{aligned}
\textbf{Proposed activation cost} &=\sum_{i=0} OH_i  \times OW_i \times O_i \times  1\\
&+\sum_{j=0}O_j \times \tfrac{4}{8} \\
&+\sum_{k=0}O_k \times \tfrac{2}{8}
\end{aligned}
\label{eq_quan_o}
\end{equation}

We compare our results with \cite{balle2018variational}, \cite{Cheng2019EnergyCI} and \cite{cheng2019deep}. In \cite{balle2018variational}, the kernel size for the main path is $5 \times 5$ which is same with ours. Besides, \cite{balle2018variational} also used hyper-prior, thus the encoder and decoder composition is also same with us. Noted that GDN is used in \cite{balle2018variational} which will actually consume some weights. Considering the weights of GDN is very few compared with the weights of the convolutional kernel, they are excluded from the weight consumption comparison. Compared with \cite{balle2018variational}, we can save more than 70\% weight and activations for both encoder and decoder. For \cite{Cheng2019EnergyCI}, there is no hyper path thus the encoder is composed of $g_a$ and the decoder is composed of $g_s$, respectively. Compared with \cite{Cheng2019EnergyCI}, we can save more than 90\% memory cost for the activation. About the weight, our encoder consumes 4.48\% more weight consumption since there is no hyper path in \cite{Cheng2019EnergyCI}. About the decoder, we can reduce about 21\% weight storage. Compared with \cite{cheng2019deep}, we can save around 75\% memory cost for weight and activation.

{

\subsection{Computational Cost Evaluation}

By using the fixed-point quantization, each weight and activation element is quantized from 32-bit floating point to 8-bit fixed point. As reported in \cite{isscc}, the energy cost is significantly reduced when using 8-bit integer compared with 32-bit floating point. The energy cost evaluation is shown in Table \ref{table_energy}. We can see that for the multiplication, 32-bit floating point consumes 3.7pJ, while 8-bit integer only consumes 0.2pJ. For the addition, 32-bit floating point consumes 0.9pJ, while 8-bit integer only consumes 0.03pJ. In the real implementation, the fixed-point calculation can be operated as the integer calculation with shifting. As a result, the computational cost can be significantly reduced by our method.

}

\begin{table}[t]
	\setlength{\abovecaptionskip}{-0.5cm}   
	\setlength{\belowcaptionskip}{0.0cm}
	\caption{Energy costs for various operations in $pJ$}
	\footnotesize
	\begin{center}
		{
			\resizebox{0.9\width}{!}{
				
				\begin{tabular}{|c|c|c|c|c|c|}
					\hline			
					\multicolumn{3}{|c}{ Integer }
					&\multicolumn{3}{|c|}{ Floating-point } \\ \hline
					
					\text{bit-depth}					
					&\text{Add}
					&\text{Mult}
					&\text{bit-depth}				
					&\text{Add}
					&\text{Mult}						
					\\ \hline								
					
					\text{8-bit}			
					&\text{0.03}		
					&\text{0.2} 
					&\text{16-bit}		
					&\text{0.4}		
					&\text{1.1} 
					\\ \hline
					
					\text{32-bit}			
					&\text{0.1}		
					&\text{3.1} 
					&\text{32-bit}		
					&\text{0.9}		
					&\text{3.7} 
					\\ \hline
					
				\end{tabular}
				
			}
			
		}
	\end{center}
	\label{table_energy}
	\vspace{-4mm}
\end{table}

\section{Conclusion}
This paper proposes a quantized LIC for both weight and activation. First, we explore different kinds of grouping and quantization schemes, and then determine the optimal one based on the coding gain. In addition, to alleviate the coding performance loss caused by the quantization error, a weight clipping fine tuning method is proposed. For the activation, four non-linear quantization codebooks are designed to saturate the quantization range. In addition, mean-reduced quantization is exploited for the outputs of analysis transform. The results show that the proposed quantized edition can outperform the BPG in terms of MS-SSIM. For the future work, we will design the specific hardware architectures such as FPGA and ASIC, and then evaluate the power and throughput.

\section*{Acknowledgment}

The authors would like to sincerely thank Dr. Zhengxue Cheng for the comments on the paper.

\ifCLASSOPTIONcaptionsoff
  \newpage
\fi



%


\bibliographystyle{IEEEtran}
\bibliography{strings,refs}

%








\end{document}